\documentclass[journal=jctcce,manuscript=article,layout=twocolumn]{achemso}

\pdfoutput=1

\usepackage{dcolumn}
\usepackage{bm}
\usepackage{xcolor}
\usepackage{color,soul}
\usepackage{algorithm}
\usepackage{algpseudocode}
\usepackage{url} 
\usepackage{makecell}
\usepackage{multirow}
\usepackage{booktabs}
\usepackage{subfig}
\usepackage{changes}
\usepackage[normalem]{ulem}
\usepackage{makecell}
\usepackage{amsmath}

\graphicspath{{images/}}

\newcommand{\PM}{{\texttt{\detokenize{pacemaker}} }}

\newcommand{\LMPS}{{\texttt{\detokenize{LAMMPS}} }}
\newcommand{\PACE}{{\texttt{\detokenize{PACE}} }}
\newcommand{\atomsk}{{\texttt{\detokenize{atomsk}} }}
\newcommand{\pyscal}{{\texttt{\detokenize{pyscal}} }}

\def\br{{\bm r}}

\def\vii{{\bm v}}

\def\AAb{{\bm A}}
\def\BBb{{\bm B}}
\def\ace{\varphi}

\def\cBB{c} 
\def\calF{\mathcal{F}}

\def\ord{\nu}

\newcommand {\braket}[2]{ \langle #1 | #2  \rangle }
\newcommand {\braHket}[3]{ \langle #1 | #2 | #3 \rangle }
\newcommand {\ket}[1]{| #1 \rangle}

\title{Atomic cluster expansion for quantum-accurate large-scale simulations of carbon}

\author{Minaam Qamar}

\affiliation{ICAMS, Ruhr-Universit\"at Bochum, Bochum, Germany}
\email{minaam.qamar@rub.de}
\author{Matous Mrovec} 
\email{matous.mrovec@rub.de}
\affiliation{ICAMS, Ruhr-Universit\"at Bochum, Bochum, Germany}
\author{Yury Lysogorskiy} 
\affiliation{ICAMS, Ruhr-Universit\"at Bochum, Bochum, Germany}
\author{Anton Bochkarev} 
\affiliation{ICAMS, Ruhr-Universit\"at Bochum, Bochum, Germany}
\author{Ralf Drautz} 
\email{ralf.drautz@rub.de}
\affiliation{ICAMS, Ruhr-Universit\"at Bochum, Bochum, Germany}

\date{\today}
             
\begin{document}

\begin{abstract}

We present an atomic cluster expansion (ACE) for carbon that improves over available classical and machine learning potentials.
The ACE is parameterized from an exhaustive set of important carbon structures at extended volume and energy range, computed using density functional theory (DFT). 
Rigorous validation reveals that ACE predicts accurately a broad range of properties of both crystalline and amorphous carbon phases while being several orders of magnitude more computationally efficient than available machine learning models. We demonstrate the predictive power of ACE on three distinct applications, brittle crack propagation in diamond, evolution of amorphous carbon structures at different densities and quench rates and nucleation and growth of fullerene clusters under high pressure and temperature conditions.
\end{abstract}

\maketitle

\section{Introduction}

Carbon is one of the most important elements in materials science, chemistry and biology. However, its versatile chemical bonding presents a formidable challenge for the development of accurate and transferable atomistic simulation models. The development of interatomic potentials for carbon started in the 1980's with empirical formulations of the bond order~\cite{abell1985empirical, tersoff_C_1988_PhysRevLett.61.2879} and culminated with the rigorous derivation of bond-order potentials (BOP) from the electronic structure \cite{oleinik_pettifor_prl}.  When it became possible to carry out large numbers of electronic structure calculations, mainly using density functional theory (DFT) \cite{hohenberg_kohn_DFT_orig1,kohn_sham_DFT_orig2}, machine learning (ML) potentials superseded the classical potentials since they were able to reproduce the DFT data with minimal errors. Yet the inherent dependence on the reference data and poor extrapolative capabilities limit the transferability of ML potentials and often lead to nonphysical predictions for atomic configurations not included in the training dataset.  

Emerging alternatives {to these pioneering ML potentials} are polynomial/tensorial expansions, in particular the moment tensor potentials (MTP) \cite{Shapeev16} and the atomic cluster expansion (ACE)~\cite{ACE_Drautz_PhysRevB.99.014104}. By employing a mathematically complete basis of the atomic environment \cite{Dusson22}, it was demonstrated that ML frameworks like neural networks or kernel-based Gaussian process regression are not necessary for obtaining accurate interatomic potentials. In fact, ACE was shown to be not only superior in terms of accuracy but also computational efficiency~\cite{PACE_Lysogorskiy2021}.

The difficulty of modeling carbon is reflected by the fact that the pioneering developments of Tersoff~\cite{tersoff_C_1988_PhysRevLett.61.2879} and Brenner~\cite{REBO1_PhysRevB.42.9458} followed the Finnis-Sinclair (FS) \cite{Finnis84} and embedded atom method (EAM) \cite{Daw84} potentials for metals. Different from the metallic potentials, which focus on atomic energies, Tersoff introduced an empirical expression for angularly dependent bond order \cite{Tersoff86} to model the formation of covalent directional bonds in C and Si.
The Tersoff potential as well as the family of reactive empirical bond order (REBO) potentials~\cite{REBO1_PhysRevB.42.9458, REBO2_Brenner_2002} were applied widely to study properties of crystalline, amorphous and molecular carbon structures~\cite{tersoff_application1, tersoff_application2, tersoff_application3}. Nevertheless, these early potentials had a number of limitations, such as short interaction ranges, neglect of $\pi$ bonding or missing van der Waals (vdW) interactions~\cite{Pastewka2012_BOP_for_fracture, marks1_transferability, marks2_graphitization}. Some of these deficiencies were improved in subsequent modifications, for example, by introducing screening functions for a better description of bond breaking ~\cite{Pastewka_SiC_PhysRevB.87.205410,Pastewka_SiC_PhysRevB.87.205410,rebo_S_pastewka_PhysRevB.78.161402}, or explicit terms to capture long-range dispersion forces~\cite{airebo_doi:10.1063/1.481208, AIREBO_M_doi:10.1063/1.4905549,LCBOP_PhysRevB.68.024107}.  Other successful carbon potentials, such as the environment dependent interaction potential (EDIP) \cite{edip_marks_PhysRevB.63.035401,Justo98} or ReaxFF~\cite{reax_ff_vanDuin2001, reaxff_Srinivasan2015}, followed similar strategies of employing suitable functional forms based on physical and chemical intuition, yet still in an empirical way. Rigorous derivations of the $\sigma$ and $\pi$ bond orders were eventually carried out by Pettifor and co-workers~\cite{Alinaghian94,oleinik_pettifor_prb2} based on a quantum-mechanical tight-binding model~\cite{Oleinik_pettifor_aBOP_carbon,oleinik_pettifor_prb2,oleinik_pettifor_prl,MROVEC2007230}. 

Despite their relatively short history, several ML potentials have already been developed for carbon. They can be characterized by how the local atomic environment is sensed via descriptor functions~\cite{musil2021physics}. To fulfill fundamental physical symmetries, these functions should be invariant under translation, rotation, inversion, and permutation of atoms of the same chemical species. For every atom, usually hundreds or thousands of different descriptor functions need to be evaluated such that their numerical values form the input for an ML algorithm that predicts the atomic property. The most prominent ML approaches include neural network potentials (NNP) \cite{NN_Behler_parinello_sym_PhysRevLett.98.146401} and kernel based methods, in particular the Gaussian approximation potentials (GAP) \cite{kernel_regression_GAP_PhysRevLett.104.136403}, with descriptors based on the atom centered symmetry functions (ACSF) \cite{ml_behler_C1CP21668F} or the smooth overlap of atomic positions (SOAP)~\cite{SOAP_PhysRevB.87.184115}, respectively. 

First  NNP models were tailored for specific applications, such as transformations between graphite and diamond or behavior of multilayered graphene~\cite{nnp_carbon_gradia_PhysRevB.81.100103,nnp_carbon_gradia_Khaliullin2011, NNP_Gr_PhysRevB.100.195419}. Recent NNPs aim at improved transferability by employing comprehensive reference datasets; PANNA~\cite{Shaidu2021_PaNNA2021} was built using an iterative self-consistent workflow, and DeepMD~\cite{deep_MD_carbonpot_WANG20221}  was trained on a large dataset of  bulk and low dimensional phases as well as snapshots from ab initio molecular dynamics (AIMD).  Parallel to the NNPs, several GAP parametrizations for carbon were also developed. The first GAPs focused on simulations of liquid and amorphous carbon~\cite{gap17_PhysRevB.95.094203} and pristine graphene~\cite{GAP_graphene_PhysRevB.97.054303}. In 2020, a large number of diverse carbon structures were employed to obtain a widely transferable GAP (GAP20)~\cite{gap20_doi:10.1063/5.0005084}. A closely related TurboGAP parameterization used a more efficient implementation of the SOAP descriptor~\cite{TurboGAP, soap_turbo_PhysRevB.100.024112}. These GAPs were applied to study complex phenomena, such as vapour deposition of amorphous carbon films~\cite{GAP_application_thinfilm_PhysRevB.102.174201} or the effect of defects on the corrugation of graphene~\cite{Thiemann_2021_corrugation_graphene}. Overall, a GAP was found to be the most accurate model among fourteen carbon potentials in predicting realistic amorphous structures in a recent benchmark study~\cite{marks1_transferability}.

Different from these ML potentials, which often employ empirical descriptor functions, the basis of ACE is mathematically complete. This means that ACE parameterizations can be improved and converged systematically. Furthermore, the hierarchical basis not only enables ACE to represent many other ML potentials~\cite{ACE_Drautz_PhysRevB.99.014104} but also to relate ACE to physically and chemically intuitive classical models. The physically motivated representation, which can be linear or mildly non-linear (see below), in combination with a consistent reference dataset helps to ensure that ACE asserts genuine transferability and is not plagued by the reproducibility-crisis of ML-based science \cite{Kapoor22}.

Here we present a first ACE parametrization for carbon which is not only more accurate and transferable than any of the previous potentials but also significantly more computationally efficient. We compare it in detail to the best available ML potentials for carbon and provide performance indicators for several other potentials. The excellent transferability and predictive power of ACE is highlighted on three distinct applications - brittle crack propagation in diamond,  formation of amorphous carbon structures at different quench rates, and nucleation and growth of fullerene clusters from gas phase at high pressures and temperatures.

The paper is organized as follows. In the following two sections, we provide a brief theoretical overview of the ACE formalism and elucidate how common classical potentials can be understood as simplified representations of a general ACE descriptor. In section \ref{sec:training}, we describe the training dataset and the fitting {protocol employed for construction of the carbon model. A detailed assessment} of the parametrization with respect to a test dataset to determine the general quality of the parametrization is also provided.  In section \ref{sec:validation}, we subject the potential to multiple validation tests that show the model's ability to predict structural, elastic, vibrational and thermodynamic properties of perfect bulk phases as well as defects. The ACE predictions are compared with those of the reference electronic structure calculations and {the latest} available ML carbon potentials, namely GAP20, TurboGAP and PANNA. In section~\ref{sec:applications}, we apply the model to perform three large scale simulations to investigate the brittle crack propagation in diamond, the formation of amorphous carbon structures, and the growth of fullerene molecules from the gas phase.

\section{Basics of the atomic cluster expansion}\label{sec:ace_basics}

The atomic cluster expansion provides a complete set of basis functions that span the space of local atomic environments. We summarize only the essentials of ACE here and direct interested readers to Refs.~\cite{ACE_Drautz_PhysRevB.99.014104, dusson2019atomic, drautz2020atomic, PACE_Lysogorskiy2021, pace_bochkarev_PhysRevMaterials.6.013804}.

An atomic property $p$ that is a function of the local atomic environment of atom $i$ is expanded as
\begin{equation}
\ace_i^{(p)} = \sum_{\vii} \cBB_{\vii}^{(p)} \BBb_{i \vii} \,, \label{eq:ACE}
\end{equation}
with expansion coefficients $\cBB_{\vii}^{(p)}$, and basis functions $\BBb_{i \vii}$ with multi-indices ${\vii}$. The energy of atom $i$ can then be evaluated using a linear expansion as
\begin{equation}
    E_i = \ace_i^{(1)} \,,
\end{equation}
for only one atomic property ($p=1$).  Or alternatively, if more properties are used,
\begin{equation}
E_i = \calF(\ace_i^{(1)},  \ace_i^{(2)},  \dots, \ace_i^{(P)}) \,,
\end{equation}
where $\calF$ in general is a non-linear function.
In the present ACE model, the energy is expressed using two contributions, a linear term and a square root term,
\begin{equation}
E_i =  \ace_i^{(1)} + \sqrt{\ace_i^{(2)}}  \,. \label{eq:EFS}
\end{equation}

The basis functions $\BBb_{i \vii}$ depend on atomic positions and are ordered hierarchically, which enables a systematic convergence of ACE by incrementally increasing the number of basis functions. The basis functions fulfill the fundamental translation, rotation, inversion and permutation (TRIP) invariances for the representation of scalar variables, or equivariances for the expansion of vectorial or tensorial quantities. This is achieved by taking linear combinations of basis functions, which do not necessarily fulfill any particular symmetries, as
\begin{equation}
\BBb_{i \vii} = \sum_{\vii'} \pmb{C}_{\vii \vii'} \AAb_{i \vii'} \,. 
\end{equation}
where the generalized Clebsch-Gordan coefficients $\pmb{C}$ act as a filter and remove basis functions that are not invariant under {rotation or inversion.}

For numerical efficiency the basis functions $\AAb$ are constructed recursively \cite{dusson2019atomic, PACE_Lysogorskiy2021}, and the order of the product $\ord$ determines the body order of a basis function
\begin{equation}
    {\AAb}_{i\vii} = 
    \prod_{t = 1}^{\ord} A_{i \vii_t} \,. \label{eq:AA}
\end{equation}
The atomic base $A$ is obtained by projecting local basis functions on the atomic density
\begin{equation}
    A_{i\vii} = \braket{\phi_{\vii}}{\rho_i} \,,
\end{equation}
with the atomic density centered on atom $i$
  \begin{equation}
\rho_i  = \sum_j^{j \neq i} \delta( \pmb{r} - \pmb{r}_{ji}) \,.
\end{equation} 
The local basis functions are expressed as
\begin{equation}
\phi_{\vii} = R_{nl}(r_{ji}) Y_{lm}(\hat{\br}_{ji}) \,, \label{eq:LCAO}
\end{equation}
with $r_{ji}$ being the distance from atom $i$ to $j$ that enters in the radial functions $R_{nl}$, while the spherical harmonics $Y_{lm}$ depend on the direction $\hat{\pmb{r}}$. The index $\vii = (nlm)$ is cumulative, where $n$ differentiates between orbitals with the same angular quantum number $l$ and $m$. 

The basis functions can represent local descriptor functions that are used in ML potentials as well as density and angular functions from classical potentials \cite{ACE_Drautz_PhysRevB.99.014104, drautz2020atomic}. If decomposed into explicit many-atom functions, the two-body basis functions are given by radial functions
\begin{equation}
B_{i\vii} = R_{n0} (r_{ji}) \,, \label{eq:B1exp}
\end{equation}
while the three-body terms have the form
\begin{equation}
B_{i\vii} = \frac{1 }{2l +1} \sum_{jk} R_{n_1 l} (r_{ji}) R_{n_2 l} (r_{ki}) P_l( \cos \theta_{jik}) \,,  \label{eq:B2exp}
\end{equation}
with Legendre polynomials $P_l(x)$. Expressions for higher body orders can also be obtained but are more complex~\cite{ACE_Drautz_PhysRevB.99.014104}.

\section{ACE as a generalization of classical potentials}\label{sec:ace_classical_pots}

To elucidate that ACE is not only a formally complete expansion but that it also can be considered as a systematic generalization of classical interatomic potentials, we sketch the link between ACE and the {second-moment approximation} (SMA) of electronic structure~\cite{pettifor1995bonding, finnis2003interatomic,Friedel69,Ducastelle70}. 

Starting from DFT, the energy functional can be decomposed into the band energy and  a double-counting contribution,
\begin{equation}
E^{(\text{DFT})} = E_{band} + E_{dc} \,.
\end{equation}
The band energy is given by $E_{band} = \sum_n f_n \epsilon_n$, with occupation numbers $f_n$ and eigenvalues $\epsilon_n$.  The eigenstates can then be expanded using local orbitals $\phi_{i \alpha}$ [cf. Eq.~(\ref{eq:LCAO})] as 
\begin{equation}
\psi_n = \sum_{i \alpha} \braket{i \alpha}{n} \phi_{i \alpha} \,.
\end{equation}
We assume the basis to be complete and for ease of notation consider the local orbitals to be orthonormal. 
The band energy is then represented as
\begin{equation}
E_{band} = \sum_{i \alpha} \int^{\epsilon_F} \epsilon\, n_{i\alpha}(\epsilon) \,d\epsilon = \sum_{i\alpha \, j \beta} \Theta_{i \alpha j \beta} H_{j \beta i \alpha } \,.  \label{eq:onin}
\end{equation}
The first identity is the the so-called onsite representation of the band energy, where the local density of states $n_{i\alpha}(\epsilon)$ of orbital $\alpha$ on atom $i$ is filled with electrons up to the Fermi level $\epsilon_F$.\footnote{For a discrete spectrum the local density of states is given as $n_{i\alpha}(\epsilon) = \sum_n |\braket{i \alpha}{n}|^2 \delta(\epsilon - \epsilon_n)$.}
The second identity, the so-called intersite representation, involves the density matrix/bond order
\begin{equation}
\Theta_{i \alpha j \beta} = \sum_n f_n \braket{i \alpha}{n} \braket{n}{j \beta} \,,
\end{equation}
and Hamiltonian matrix elements $H_{i \alpha j \beta} = \braHket{i \alpha}{\hat{H}}{j \beta}$. 
A formal expansion of the DFT functional with respect to charge density \cite{harris1985simplified, sutton1988tight, Foulkes89, frauenheim2000self, finnis2003interatomic, Drautz_PhysRevB.84.214114_2011, Drautz15} presents the basis of modern {tight-binding} (TB) models and results in a partitioning of the energy as
\begin{equation}
E^{(\text{TB)}} = E_{bond} + E_{prom} + E_{rep} \,,
\end{equation}
where for simplicity we neglected charge transfer. If there is no promotion of electrons and the energy scale is set such that $H_{i \alpha i \alpha} = 0$, the band and bond energies are identical, $E_{band} = E_{bond}$. The repulsive energy $E_{rep}$, comprising the double-counting contribution and Coulomb interactions between the atomic cores, is often approximated by a pair potential
\begin{equation}
E_{rep} = \frac{1}{2} \sum_{ij} V(r_{ij})\,.
\end{equation}
For the derivation of the {second-moment approximation} for metals, we utilize a local expansion of the band energy for an atom 
\begin{equation}
E_{band,i} = \int^{\epsilon_F} \epsilon\, n_{i}(\epsilon) \,d\epsilon \,\,\,\,\, \text{with} \,\,\,\,\, n_{i}(\epsilon)  = \sum_{\alpha} n_{i \alpha}(\epsilon) \,, 
\end{equation}
We next construct the local density of states from the information about the local atomic environment. This is achieved by the recursion method~\cite{Ducastelle70,Friedel69,Haydock80}. If the recursion is continued with constant coefficients after the first recursion level, only information up to the second moment of the density of states enters the expansion. The second moment, given by
\begin{align}
\mu_i^{(2)} &= \int \epsilon^2 n_{i}(\epsilon) \,d\epsilon = \sum_{\alpha\, j \beta} H_{i \alpha j \beta} H_{ j \beta i \alpha} \, , 
\end{align}
is determined by Hamiltonian matrix elements that rapidly decay with increasing distance between atoms $j$ and $i$. As the zeroth moment $\mu_i^{(0)}=1$ and the first moment was set to $\mu_i^{(1)} =  \sum_{\alpha} H_{i \alpha i \alpha} = 0$, it can be viewed merely as a consequence of appropriate scaling that in {second-moment approximation}~\cite{Gupta81, pettifor1995bonding, Ackland_1988}
\begin{equation}
E_{band,i}^{(\text{SMA})} = C \sqrt{\mu_i^{(2)} }  \propto \sqrt{\mathcal{Z}_{i}} \,, \label{eq:2Mband}
\end{equation}
where the pre-factor $C$ is a function of band filling and $\mathcal{Z}_{i}$ is the coordination, the number of nearest neighbors, of atom $i$. The total atomic energy can then be written as 
\begin{equation}
E_{i}^{(\text{SMA})} = C \sqrt{\mu_i^{(2)} } + \sum_j V(r_{ij})\,.
\end{equation}
As $\mu_i^{(2)}$ is strictly positive, it may also be understood as an atomic density $\rho_i$ computed from pairwise functions to neighboring atoms. In this way, one obtains immediately the Finnis-Sinclair potential~\cite{Finnis84}.  If the square root function is replaced by a general, concave embedding function one arrives at the EAM formulation~\cite{Daw84}. Therefore, the cohesion in metals in {second-moment approximation} does not increase linearly with increasing coordination and correctly reflects the unsaturated nature of the metallic bond~\cite{pettifor1995bonding, heine1991many}.

The derivation of the {second-moment approximation} expressions for covalent elements is somewhat more involved and requires explicit consideration of the angular character of atomic orbitals. By equivalence of the onsite and intersite representations of the band energy [cf. Eq.~(\ref{eq:onin})], second-moment approximation for the band energy also implies a corresponding expression for the bond order. We assume the Slater-Koster two-center approximation \cite{Slater54} with the $z$-axis of the coordinate system aligned along the bond $i -j$.  For a d-valent atom, the Hamiltonian matrix is diagonal with the matrix elements equal to two-center distance-dependent bond integrals dd$\sigma(r_{ij})$, dd$\pi(r_{ij})$ and dd$\delta(r_{ij})$.
The second moment is then by construction invariant under rotation and given by
\begin{equation}
 \mu_i^{(2)} = \sum_{j} \left[ \mathrm{dd}\sigma(r_{ij})^2 + 2\, \mathrm{dd}\pi(r_{ij})^2 + 2\, \mathrm{dd}\delta(r_{ij})^2 \right] \,, \label{eq:dSMA}
\end{equation}
where the summation is over the $\mathcal{Z}_{i}$ neighbors of atom $i$. 
The $\sigma$ bond order for the $i-j$ bond (with analogous expressions for the $\pi$ and $\delta$ bond orders) is expressed as~\cite{pettifor1995bonding, finnis2003interatomic}
\begin{equation}
| \Theta_{ij}^{(\sigma)} | = 
\frac{C }{ \sqrt{ \mu_i^{(2)}  } } \propto \frac{1}{\sqrt{\mathcal{Z}_{i} }}\,, \label{eq:BO}
\end{equation}
and is thus inversely proportional to the square root of the number of neighbors of atoms $i$. Note that the bond order is not symmetric with respect to exchange of atoms $i$ and $j$ and therefore the denominator needs to be replaced by $\sqrt{( \mu_i^{(2)} + \mu_j^{(2)})/2}$.

Next, since the bond (or band) energy of atom $i$ 
\begin{equation}
E_{bond,i}^{(\sigma)} =  \sum_j \Theta_{ij}^{(\sigma)} dd\sigma(r_{ij}) \propto \sqrt{\mathcal{Z}_{i}}  \,. \label{eq:ebond_dds}
\end{equation}
is obtained as the sum over all $\mathcal{Z}_{i}$ neighbors, it attains the square root dependence with the number of neighbors as in Eq.~(\ref{eq:2Mband}).

The derivation for sp-valent elements follows along the same lines but needs to take into account the directionality of the hybrid orbitals and the energy splitting $\Delta E_{sp}$ of the s and p orbitals. The hybrid $\sigma$ orbitals oriented along the $z$-axis are formed as
\begin{equation}
\ket{i\sigma} = \frac{1}{\sqrt{1+\lambda^2}} \left ( \ket{is} + \lambda \, \ket{i p_z} \right ) \,,
\end{equation}
where $\lambda=1$, $\sqrt{2}$ and $\sqrt{3}$ correspond to sp, sp$^2$ and sp$^3$ hybrids, respectively. The bond integral of the $\sigma$ hybrid is given as
\begin{equation}
h_{\sigma}(r_{ij}) = \frac{ \mathrm{ss}\sigma(r_{ij}) - 2 \lambda\, \mathrm{sp}\sigma(r_{ij}) - \lambda^2 \, \mathrm{pp}\sigma(r_{ij}) }{1+\lambda^2}  \, ,
\end{equation}
where ss$\sigma$, pp$\sigma$ and sp$\sigma$ are Slater-Koster two-center bond integrals.
The second moment takes the form \cite{pettifor1995bonding}
\begin{equation}
\begin{split}
 \mu^{(2)}_{i\to j} &=  c \, \Delta E_{sp}^2 + h_\sigma(r_{ij})^2 + \\
 & \sum_{k \ne i,j} \frac{1}{2} \left(  h_\sigma(r_{ik})^2 \, g(\theta_{jik}) + h_\sigma(r_{jk})^2 \, g(\theta_{ijk}) \right)  \,.
\end{split}
\end{equation}
which differs from Eq.~(\ref{eq:ebond_dds}) due to the angular functions $g_{jik}$ and $g_{ijk}$ which depend on the angle $\theta$ between bonds~\cite{pettifor1995bonding, Alinaghian94}.  Different from the d-valent case, the second moment is not rotationally invariant and the orientation of the $z$-axis along the bond $i \to j$ has to be given explicitly. 
The bond order retains a form analogous to Eq.~(\ref{eq:BO}),
\begin{equation}
 |\Theta_{i \to j}^{(\sigma)} | = \frac{C}{ \sqrt{ \mu_{i\to j}^{(2)} } } \,.
\end{equation}
This expression for the bond order is not dissimilar from the empirical bond order introduced by Tersoff \cite{Tersoff86}. Derivations of the $\pi$ bond order and the promotion energy, which are important for the bond formation in carbon, can be found in Refs.~\cite{Alinaghian94,Oleinik_pettifor_aBOP_carbon, pettifor1995bonding, finnis2003interatomic}. 

The aim of this analysis was to show that the most important classical potentials for metals and covalent semiconductors can be understood from the {second-moment approximation}. The crucial point is that for both these materials, {second-moment approximation} predicts the bond energy to scale as a square root of the local atomic density. This suggests that a physically based model of the atomic energy should comprise an attractive part with a square-root dependence on the number of neighbors, and a pair-wise repulsive part which scales linearly with the number of neighbors. This expression for the energy,
\begin{equation}
E_i = - A \sqrt{\mathcal{Z}_i} + B \mathcal{Z}_i \,
\end{equation}
mimics the ACE formulation in Eq.~(\ref{eq:EFS}). If we limit ACE to two-body basis functions, the representation of the energy is closely related to Finnis-Sinclair models. If 2-body and 3-body contributions are included in the ACE basis, we expect ACE to reproduce not only the original Tersoff formulations but also Stillinger-Weber \cite{Stillinger85}, EDIP \cite{Justo98} and other empirical bond order potentials. Incorporation of basis functions with higher body orders then represents a systematic generalization. The higher body orders are similar to including higher moments of the density of states as required for more accurate structural differentiation~\cite{pettifor1995bonding}. An ACE parametrization with contributions up to the body order of six was developed recently for metallic Cu \cite{PACE_Lysogorskiy2021}. Here we show that this approach also works extremely well for covalent carbon, where directional bonding is much more important and delicate.

\section{Training of the ACE for carbon}
\label{sec:training}

\begin{figure*}
\centering
    \includegraphics[width=0.99\textwidth]{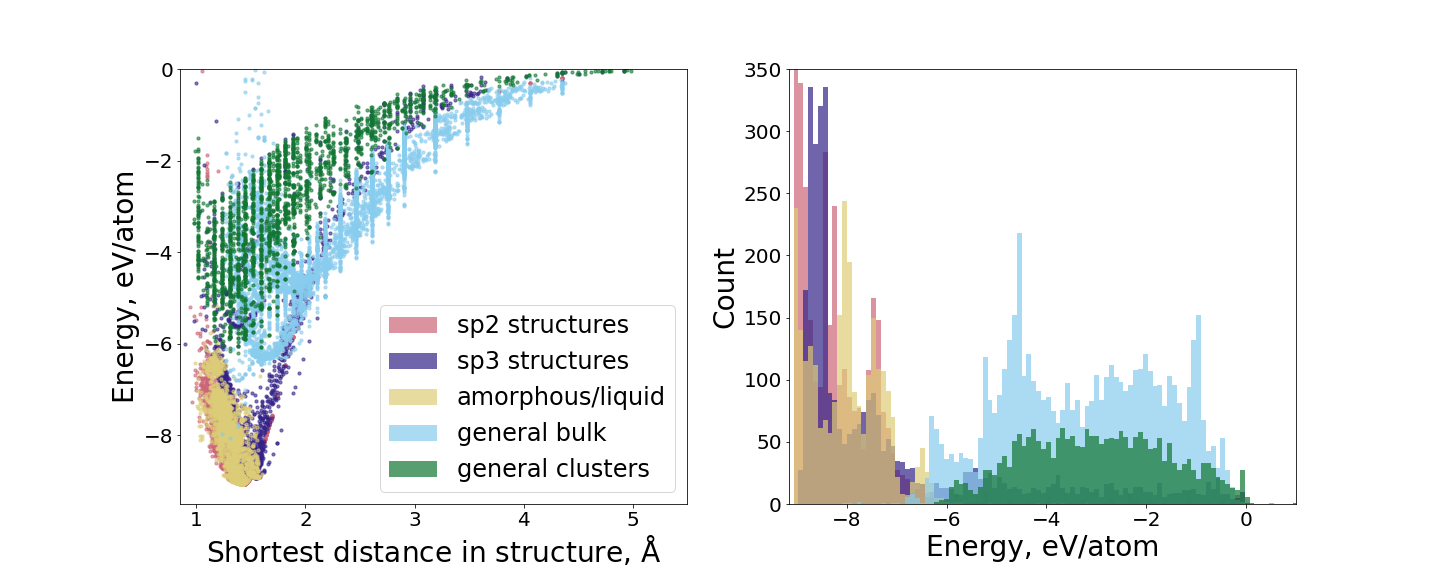} 
    \caption{\label{fig:dataset} Training dataset visualized as a scatter plot of energy per atom with respect to the nearest interatomic distance within each structure and (b) the distribution of cohesive energies.}
\end{figure*}

An accurate and consistent reference dataset that covers a large part of the phase space of atomic configurations is critical for the construction of an ACE model. Such a dataset consists of a series of atomic structures and their corresponding energies, forces and stresses, typically evaluated using electronic structure methods such as DFT. In this section, we describe the details of our DFT calculations including peculiarities related to carbon, present our strategy to generate an exhaustive and balanced set of reference structures, and show how the ACE parametrization is carried out.

\subsection{DFT reference and dispersion interactions}
\label{sec:dft_settings}

The DFT reference calculations were performed using the Vienna Ab-initio Simulation Package (VASP)~\cite{dft1,dft2,dft3}, version 5.4.4. The exchange-correlation energy was computed using the Perdew-Burke-Ernzerhof (PBE) generalized gradient approximation (GGA)~\cite{gga} and the core electrons were modeled by the Projector-Augmented Wave (PAW)~\cite{paw1,paw2} method (C: s2p2).  We carried out highly converged calculations with tight settings of the principal parameters in order to obtain accurate results for the energy and forces. Specifically, the energy cutoff for the plane-wave basis was set to 500 eV and the convergence threshold for the energy to 10$^{-6}$ eV.  Gaussian smearing with a width of 0.1 eV was applied.
For periodic structures, the Brillouin zone was sampled using a dense $\Gamma$-centered k-point mesh with the spacing between the k-point of 0.125  
per \AA{}$^{-1}$ while for non-periodic clusters only the $\Gamma$ point was used. An additional support grid was employed for the {calculation of forces help reduce the noise}.

While the PBE functional describes accurately covalent bonds, it does not capture long-range dispersion interactions. As the vdW interaction plays a crucial role in the stabilization of many important carbon structures, such as graphite and its derivatives, PBE data alone is not suitable for the parametrization of a fully transferable carbon model. There exist various approaches to account for dispersion interactions within DFT~\cite{vdw_d2_coorection, vdw_d3_coorection,vdw_d4_coorection,vdw_TS_correction, vdw_TS_MBD, vdw_TS_MBD2}. We decided to employ additive corrections, which allow us to parameterize ACE based on standard PBE data, which is significantly shorter ranged than the vdW interactions, and then to amend the ACE model with a correction term in analogy to most dispersion-corrected DFT approaches~\cite{grimme2011density}. This approach not only results in an efficient model with the correct description at long interatomic distances, but it also gives the flexibility to employ correction terms of different complexity or even switch off the long-range interactions when required.

Finally, ACE is parametrized such that the interaction between C atoms approaches zero at infinite separation to ensure that the fitted energies correspond to cohesive energies. This is done by taking the energy of an isolated spin-unpolarized atom as the reference zero energy.

\begin{table*}
\scriptsize
\centering
\begin{tabular}{lccccc}
\hline
            Category  &  Description &  \makecell{Number\\of structures} &  \makecell{Number\\ of atoms} & \makecell{[$E_{min}$, $E_{max}$]\\ (eV/atom)} &  \makecell{NNB range\\ (\AA)} \\
\hline
      sp$^2$ structures &  \makecell{graphene, graphite, fullerenes, nanotubes,\\ incl. defects} &     3532 &  88 358 & [-9.07, 78.50] & [0.7, 4.4] \\[5pt] \hline
      sp$^3$ structures &  \makecell{cubic and hexagonal diamond, \\ high-pressure phases (bc8, st12, m32, etc.),\\ incl. defects} &     3407 &  84 290 & [-8.93, 36.99] & [0.9, 4.9] \\[5pt] \hline
    amorphous/liquid &  \makecell{selected from available datasets;\\ amorphous and liquid phases \cite{gap20_doi:10.1063/5.0005084} \\ MD trajectories of multilayered graphene \cite{NNP_Gr_PhysRevB.100.195419}} &     2642 &  146 188 & [-9.06, -3.18] & [1.0, 1.7] \\[5pt] \hline
general bulk & \makecell{basic crystals; fcc, hcp, bcc, sc, A15, etc.\\ over broad range of volume and \\random displacements of atoms/cell deformations }   &     5 342 &   39 126 & [-8.06, 82.17] & [0.9, 4.4] \\[5pt] \hline
general clusters & \makecell{non-periodic clusters with 2-6 atoms} &     2370 &    8 801 & [-6.19, 83.28] & [0.6, 5.0] \\ 
\hline
\end{tabular}
\caption{A description of the training dataset used in this work.}\label{tab:dataset_description}
\end{table*}

\subsection{Reference dataset}
\label{sec:dataset}

We constructed an extensive reference dataset consisting of 17,293 structures with a total number of 366,763 atoms. The reference structures were chosen to sample a broad range of atomic configurations for carbon to ensure good transferability of the ACE parametrization.

We divided the reference structures into five categories. The first category contains carbon structures with prototypical sp$^2$ bonding, including a variety of bulk graphite structures, 2D graphene sheets, molecular fullerenes and nanotubes. The second category contains sp$^3$ four-fold coordinated crystalline diamond structures and their high pressure variants. In both categories, we sampled the structures over a broad range of interatomic distances, shape distortions, random displacements of atoms, and incorporated point and planar defects such as vacancies, Stone-Wales defects, surfaces, interfaces, etc. These two categories were complemented by structures from MD simulations of amorphous and liquid carbon~\cite{gap20_doi:10.1063/5.0005084} and  multilayered graphene~\cite{NNP_Gr_PhysRevB.100.195419}. As the MD structures are highly correlated, we included a relatively small number of atomic configurations from these two collections.  
The last two categories, referred to as `general bulk' and `general clusters', consist of general crystal structures, e.g., fcc, bcc, sc, hcp, A15, and isolated random clusters containing up to six carbon atoms. These two categories serve to sample a broader region of configurational space and allow the potential to model close-packed atomic environments. A summary of the five categories is provided in Table~\ref{tab:dataset_description}. {From the total number of atoms, in most cases each atom has a unique environment and provides important information for the ACE fit.}

Figure~\ref{fig:dataset} shows the cohesive energy for the reference structures as a function of the shortest bond length within each structure (left panel) and the distribution of structures within the corresponding energy range (right panel). The standard PBE functional without any vdW correction predicts the graphene phase to have the lowest cohesive energy of $-9.12$~eV/atom and an equilibrium bond length of 1.45~\AA{}. The energies of most structures from the first three categories lie within 3 eV/atom above the ground state energy, while the remaining two categories are characterized by cohesive energies mostly greater than $-6$~eV/atom.  

\subsection{ACE implementation and efficiency}
\label{sec:software}

\begin{figure}
  \centering
  \includegraphics[width=0.99\columnwidth]{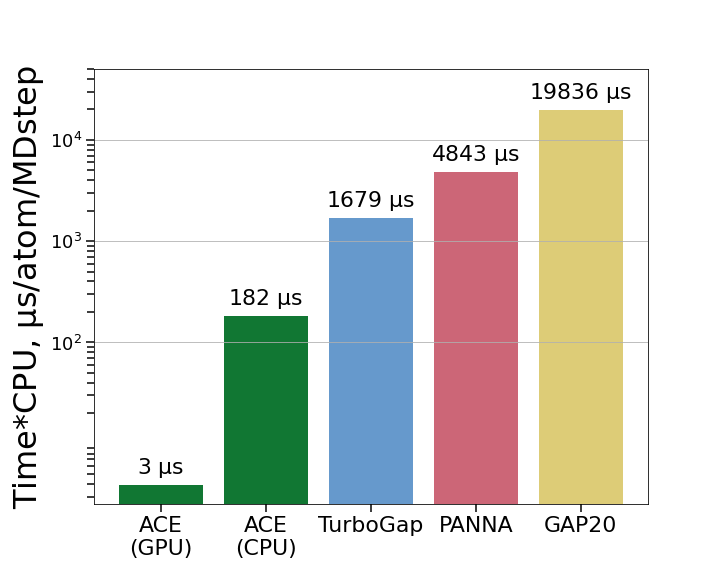}
  \caption{
    A comparison of typical CPU (AMD Ryzen 5 3600X) and GPU (Tesla V100S-PCIE 32 GB) times for ACE, TurboGAP, PANNA and GAP20 (for TurboGAP the value was taken from Ref.~\cite{TurboGAP}). CPU times are given per core and GPU times are per {device}.}
  \label{fig:walltime}
\end{figure}

For the parameterization of ACE we employed the software package \PM \cite{pace_bochkarev_PhysRevMaterials.6.013804}. The simulations for validation and applications were carried out using \LMPS \cite{LAMMPS} with the \PACE package \cite{PACE_Lysogorskiy2021}. Both \PM and \LMPS+ \PACE can be executed on CPU and GPU architectures. The computational performance of ACE in comparison with other ML methods is shown in Fig.~\ref{fig:walltime}. The graph displays the CPU/GPU times for a single MD time step per atom from representative $NVT$ MD simulations of liquid carbon at 4000 K with density of 2 g/cm$^{3}$ (using a periodic supercell containing 1000 atoms for 1000 time steps). For TurboGAP we used the value reported in Ref.~\cite{TurboGAP}, since it is not available in \LMPS.  On CPU, ACE is more than 1-2 orders of magnitude faster than the other models in accordance with our previous benchmarks~\cite{PACE_Lysogorskiy2021}. On GPU, ACE reaches efficiency comparable to that of classical interatomic potentials.

\subsection{Training}
\label{sec:fitting_process}

Training a carbon potential for an optimal balance between accuracy and transferability is more challenging than most other elemental systems. The potential must simultaneously reproduce minute energy differences between the most stable phases and allotropes while being able to capture large energy changes associated with breaking and re-arrangement of the strong directional covalent bonds that occur during phase transformations or in the vicinity of structural defects.

For ACE training, we employed a hierarchical optimization strategy implemented in the \PM code that sequentially adds basis functions in predefined steps. The radial basis functions represented by exponentially-scaled Chebshev polynomials were also included in the optimization. Structures with low cohesive energies (3 eV above the ground state) were assigned higher weights in the loss function. The employed training dataset together with the input for the \PM code is provided in the supplementary material~\cite{suppl}.

\begin{figure}
\centering
    \subfloat[]{\includegraphics[width=0.25\textwidth]{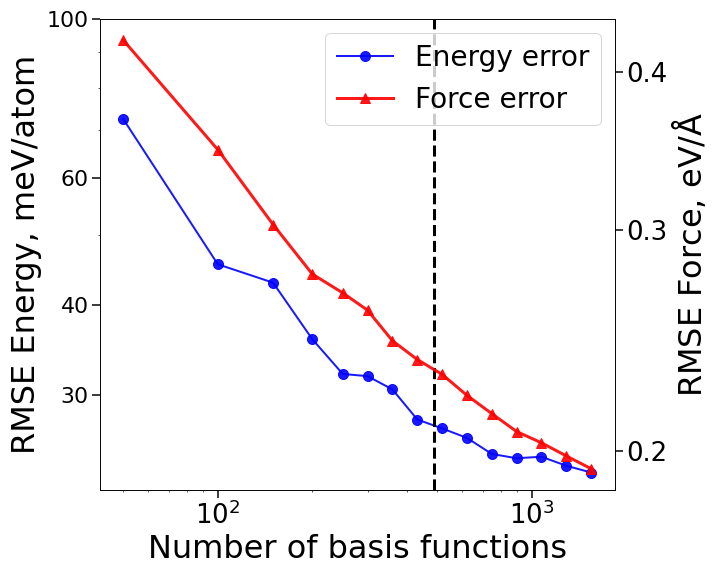}}
    \subfloat[]{\includegraphics[width=0.25\textwidth]{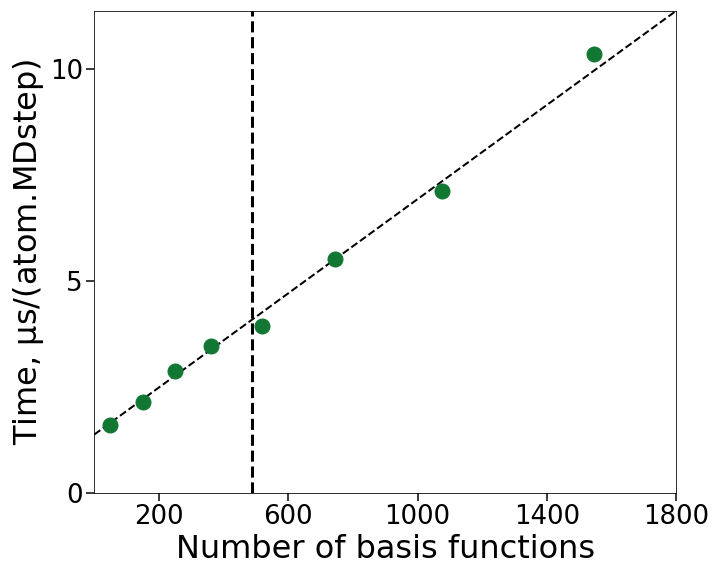}} 
    \caption{(a) The feature curves showing the RMSE of energy and force as a function of the number of basis functions; (b) the scaling of ACE computational time on GPU with the number of basis functions; vertical dashed lines mark the presented ACE parametrization with 488 basis functions.}
    \label{fig:timing_n_funcs} 
\end{figure}

The presented ACE parametrization comprises of 488 basis functions containing terms up to the fifth body order, which is sufficient to attain an outstanding overall accuracy while remaining computationally efficient. As shown {for the test fit feature curve} in Fig.~\ref{fig:timing_n_funcs}(a), the accuracy can be further improved by increasing the number of basis functions~\cite{pace_bochkarev_PhysRevMaterials.6.013804}, while the scaling of the computational time, shown in Fig.~\ref{fig:timing_n_funcs}(b), remains linear~\cite{PACE_Lysogorskiy2021}. The breakdown of RMSEs for different categories predicted by the largest potential with 1950 functions is given in Table S2 in the supplementary material~\cite{suppl}.

\begin{table}
\scriptsize
\centering
\begin{tabular}{lrrrr}
\hline
    Category (train/test) & $E_\text{train}$ & $E_\text{test}$ & $F_\text{train}$ &  $F_\text{test}$ \\
\hline
   sp$^2$ structures (3532/385) &        29 &   28 &   465 &  499 \\
   sp$^3$ structures (3407/378) &        59 &   74 &   307 &  367 \\
   amorphous/liquid (2642/315) &        56 &   54 &   587 &  567 \\
   general bulk (5342/606)      &        97 &  114 &  1332 &  1420 \\
   general clusters (2370/246)  &       154 &  186 &  1120 &  1195 \\
\hline
\end{tabular}
\caption[fit stats]{Energy and force RMSE for each category of the reference dataset. Numbers in brackets correspond to the number of structures for training and testing, respectively. Energies are in meV/atom and forces in meV/\AA{}. \label{tab:rmse_dataset}}
\end{table}

The root mean square errors (RMSE) for the different subsets are given in Table~\ref{tab:rmse_dataset}. The fact that RMSE for train and test sets are comparable demonstrates that the model is not overfitted. The optimized potential has an energy RMSE of 21 meV/atom for the structures within 3 eV/atom from the ground state and 166 meV/atom for the complete dataset.  The corresponding force RMSE amount to 218 and 689 meV/\AA{}, respectively. Most large force errors arise from high energy structures, specially from the `general bulk' and `general clusters' categories {(see Figure S4 in the supplementary material}~\cite{suppl}) and structures with short interatomic distances, which contain forces up to 100 eV/\AA{}. As reported in the supplementary material~\cite{suppl}, previous carbon models exhibit even larger errors for both forces and energies {(see Table S1 and Figure S5)}.

\begin{figure}
\centering
    
    \begin{tabular}{@{}c@{}}
    \includegraphics[width=0.99\columnwidth]{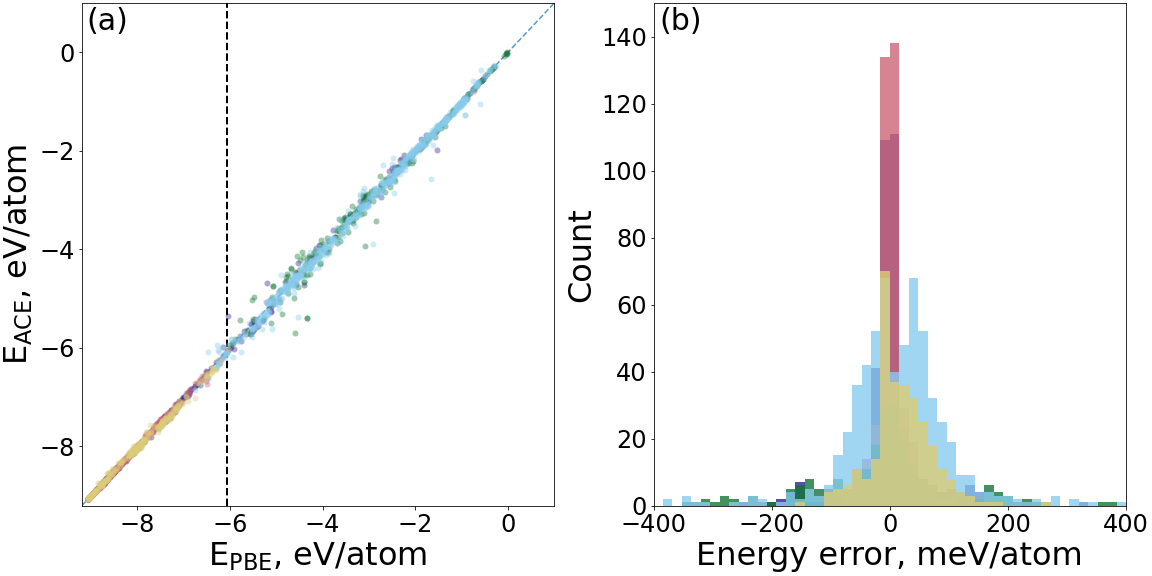}
  \end{tabular}
  
  \vspace{\floatsep}

  \begin{tabular}{@{}c@{}}
    \includegraphics[width=0.99\columnwidth]{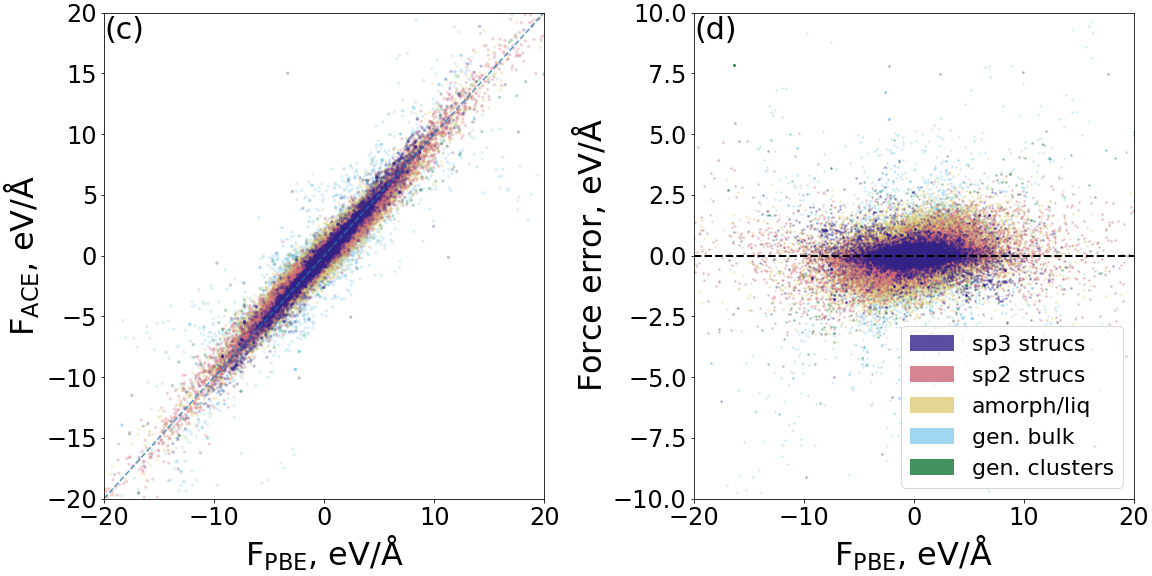}
  \end{tabular}

    \caption{\label{fig:crossref} Energy and force errors for a split test dataset consisting of 1,912 structures; the dashed vertical line in (a) marks the boundary of differently weighted data.}
\end{figure}

An overall assessment of the accuracy of presented ACE parametrization is given in Fig.~\ref{fig:crossref}. Figure~\ref{fig:crossref}(a) shows the predicted ACE energies with respect to the reference PBE data for a 10\% split test set with 1,912 structures. The effect of higher weighting of the low energy structures is clearly visible. Most structures with energies below $-6.2$ eV/atom, indicated by the vertical line, match the reference very clesely. For structures with higher energies (and lower weights), the deviations are larger. Figure~\ref{fig:crossref}(b) details the distribution of energy errors for the different categories. As these are not normalized distributions, the area under the peaks corresponds to the total number of structures within each category. The standard deviations associated with each category are considerably different, with the sp$^2$ and sp$^3$ structures having the most narrow distributions. The force cross-correlation and error distribution are plotted for all categories in Figs.~\ref{fig:crossref}(c) and (d), respectively.

\begin{figure}
\centering
\includegraphics[width=0.45\textwidth]{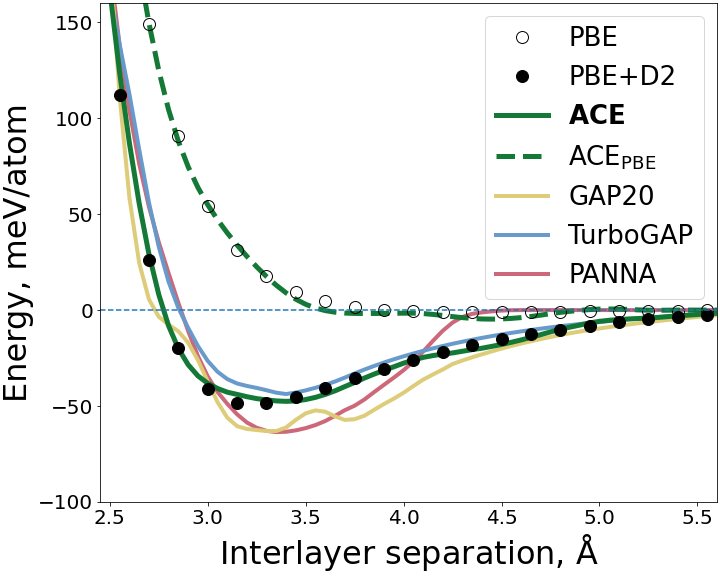}
\caption{Energy as a function of interlayer separation in graphene. {Results for other dispersion corrections and the other models are shown in the supplementary material}~\cite{suppl}.}
\label{fig:graphite_eb}
\end{figure}

Figure~\ref{fig:graphite_eb} shows the binding energy between the layers of graphite as a function of the interlayer separation. The standard PBE functional gives a negligible binding energy of 1 meV/atom at around 4.5 \AA{}, i.e., essentially only a short range repulsive interactions between the graphene sheets. This PBE reference is reproduced accurately by the base ACE (referred to as ACE$_\text{PBE}$), that was trained on the uncorrected PBE data.  By adding the D2 correction with a long range cutoff of 9 \AA{}, ACE achieves an excellent description of cohesion in graphite, very close to that of TurboGAP. While qualitatively similar, GAP20 shows rather oscillatory behavior whereas for PANNA the range of vdW interactions is significantly underestimated. {Additional results for other dispersion corrections and other NNPs are given in Figs. S1(a) and S2(a) in the supplementary material}~\cite{suppl}.

\section{Validation}
\label{sec:validation}

We carried out multiple validation tests to assess the performance of the carbon ACE. In the following we present key tests and compare the predictions of ACE to those of the best available potentials. Further validations are provided in the supplementary material~\cite{suppl}.

\subsection{Stability of bulk phases} \label{sec:basic_tests}

\begin{figure}
\centering
\includegraphics[width=0.45\textwidth]{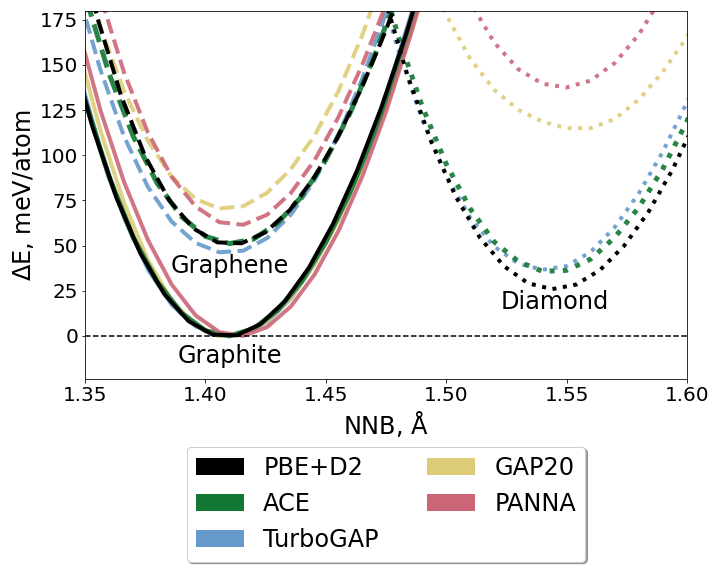}
\caption{Relative stability of graphite, graphene and diamond as predicted by PBE+D2, ACE and the best available models. $\Delta E$ with respect to graphite as given by the respective potential. Result for other NNPs are presented~\cite{suppl}.}
\label{fig:ev_groundstate}
\end{figure}

One of the peculiarities of bonding in carbon is that the energy of three-fold coordinated graphite and graphene is nearly the same as that of four-fold coordinated diamond, while the energy of the carbon dimer is much higher. This illustrates the importance of strong angular bond contributions as well as weak dispersion interactions. According to our PBE+D2 calculations, the graphite ground state is separated from those of diamond and graphene by only 29 and 50~meV/atom, respectively. This energy ordering is in agreement with experimental~\cite{wagman1945heats} as well as recent theoretical predictions using a high-level coupled cluster theory, according to which diamond lies less than 30 meV/atom above graphite~\cite{Popov_2019_C8CP07592A, Gruber_PhysRevX.8.021043}. These subtle energy differences are, however, not captured correctly by all DFT functionals.  While our PBE+D2 results agree well with those of the PBE+MBD \cite{mbd1_PhysRevLett.108.236402,mbd2_doi:10.1063/1.4865104} functional, which was used to generate the TurboGAP reference data, the hybrid  optB88-vdW \cite{opt1_PhysRevB.82.081101,opt2_PhysRevB.83.195131,opt3_PhysRevLett.92.246401,opt4_Klime__2009} and rVV10 \cite{rVV10_PhysRevB.87.041108} functionals, which were used for the GAP20 and PANNA datasets, respectively, predict diamond to have a significantly larger energy than both graphite and graphene. The consequence of different training data is illustrated in Fig.~\ref{fig:ev_groundstate}, which shows the relative stability of the three structures as a function of nearest-neighbor (NNB) distance obtained by ACE, TurboGAP, GAP20 and PANNA. While ACE and TurboGAP match closely the PBE+D2 reference data, GAP20 and PANNA predict that diamond is substantially less stable than both sp$^2$ allotropes.

\begin{figure}
\centering
    \includegraphics[width=0.5\textwidth]{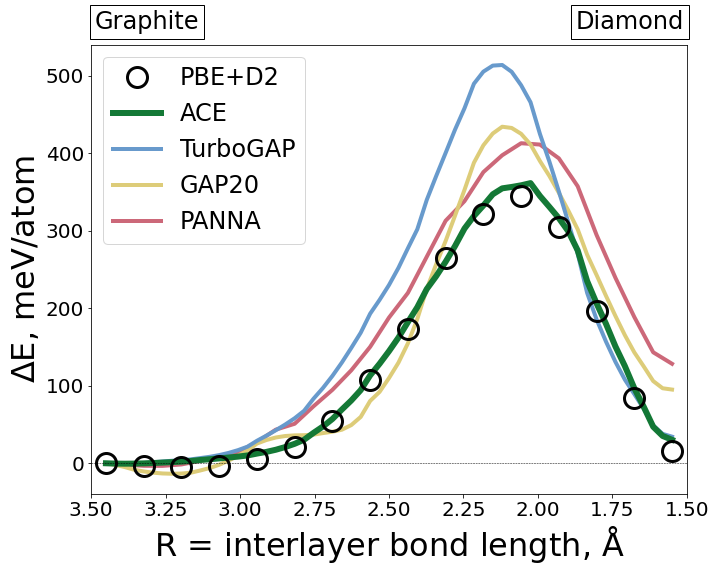}
    \caption{Energy barrier associated with the graphite to diamond transformation; $\Delta E$ with respect to graphite. \label{fig:gradia}
    }
\end{figure}

Figure~\ref{fig:gradia} shows the energy barrier associated with the transformation between AB-stacked rhombohedral graphite and diamond, which proceeds by simultaneous buckling and lateral compression of graphene sheets~\cite{gradia_transform_PhysRevB.34.1191}. According to PBE+D2, the barrier is 350 meV/atom with respect to graphite. ACE reproduces the barrier within a few meV. TurboGAP, GAP20 and PANNA overestimate it by 164, 97 and 95 meV/atom, respectively.

\begin{figure}
\centering
    \includegraphics[width=0.44\textwidth]{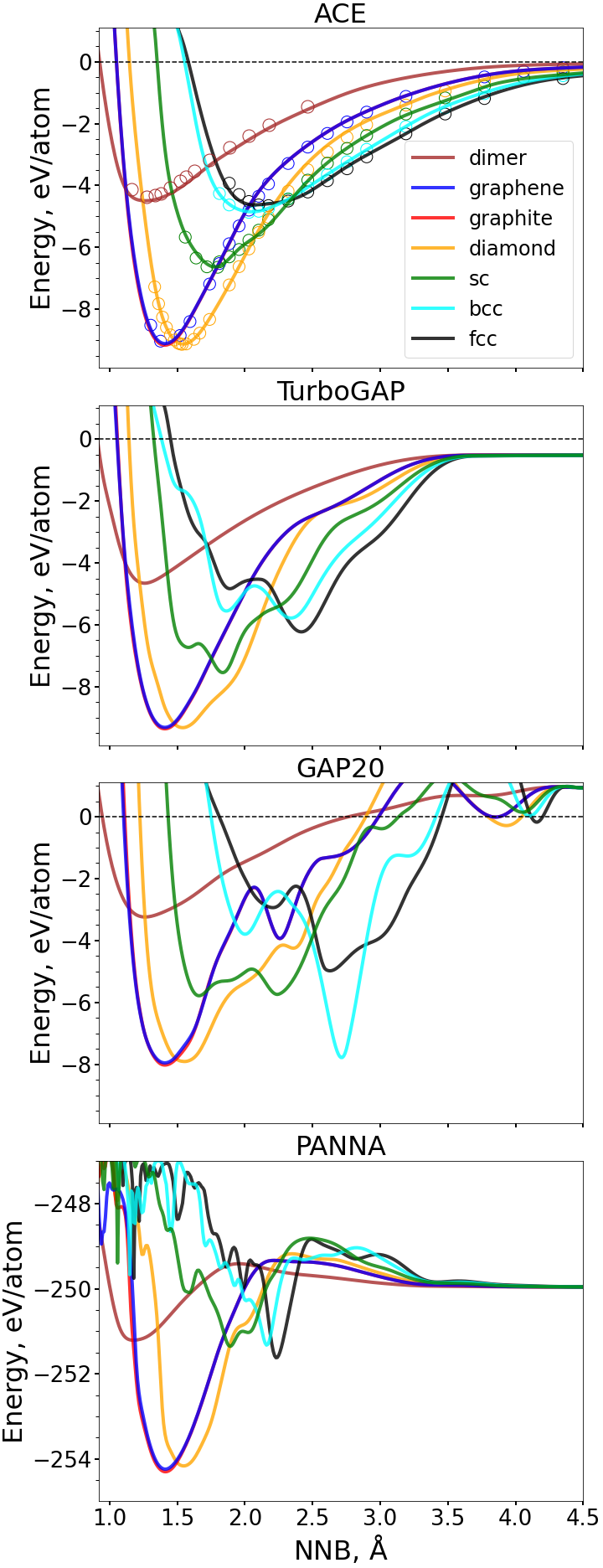}
    
    \caption{\label{fig:ev_curves} Binding energy vs NNB distance for various carbon structures predicted by ACE, TurboGAP, GAP20 and PANNA. Symbols in ACE panel represent the PBE+D2 reference. Binding energies for other carbon potentials cen be found in the supplementary material~\cite{suppl}.}
\end{figure}

In Fig.~\ref{fig:ev_curves} we compare binding energy curves for various carbon structures obtained by ACE, TurboGAP, GAP20 and PANNA.  Results for other carbon potentials are provided in {Fig. S6 in} the supplementary material~\cite{suppl}. ACE describes the binding energy curves in excellent agreement with the DFT+D2 reference over the whole range of considered NNB distances. This may be attributed to the extensive reference dataset and the extrapolation capabilities of ACE basis functions~\cite{PACE_Lysogorskiy2021}. In contrast, TurboGAP, GAP20 and PANNA reveal the well known inability of most ML potentials to extrapolate outside of the reference dataset. The three models were fitted mostly to configurations with densities close to those of equilibrium graphite and diamond, and they clearly fail to describe the low-density structures. The problem is more severe for GAP20, as it predicts large unphysical oscillations for nearly all structures beyond the NNB distance of about 1.7 \AA{}. For all structures there exist multiple local minima. These can lead to occurrence of spurious phases, for instance, in the vicinity of defects or when the system is subject to external loads. The local minima and maxima can further affect forces and undermine the description of bond making and breaking~\cite{rebo_S_pastewka_PhysRevB.78.161402}.  Finally, even though carbon does not readily form structures like fcc, bcc or sc, it is advisable to reproduce properties of these structures as well, as they may occur in MD simulations under non-equilibrium conditions or at high pressure. For example, atomistic models of amorphous carbon are often generated by melting a sc lattice of C atoms~\cite{marks1_transferability,marks2_graphitization,Jana_2019}, and the sc phase is even found to be stable at extreme pressures~\cite{Willman22,sun_2009}. 

It is worth mentioning that the energy of an isolated atom, which represents the reference for the cohesive energy, is different for the four models. For ACE, the energy of a non-magnetic free atom is subtracted from the reference data, as discussed in Sec.~\ref{sec:dft_settings}, and therefore the energy tends to zero as the atoms are pulled apart beyond the chosen cutoff. For TurboGAP, GAP20 and PANNA, this limit is not zero (cf. Fig.~\ref{fig:ev_curves}) but $-0.51$, $+0.94$ and $-249.96$ eV/atom, respectively.

\subsection{Elastic and vibrational properties} \label{sec:elc_phon}

\begin{figure}
\centering
    \includegraphics[width=0.5\textwidth]{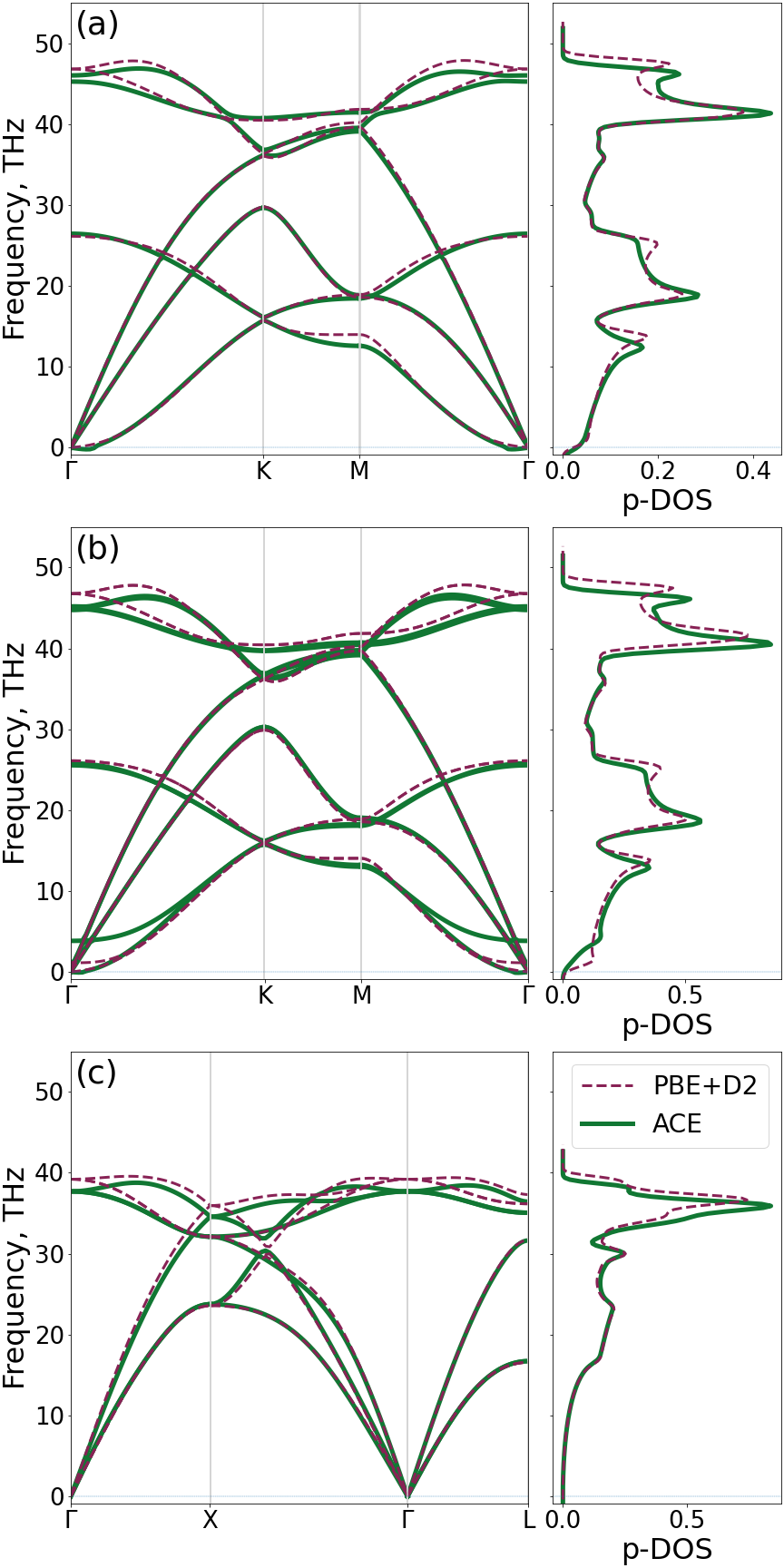}
    \caption{Phonon band structures and densities of states for (a) graphene, (b) graphite, and (c) diamond predicted by ACE (solid lines) and PBE+D2 (dashed lines). \label{fig:phonons} }
\end{figure}

\begin{table*}[]
\tiny
\begin{tabular}{l|ccc|ccc|ccc|ccc|ccc|}\hline
    & \multicolumn{3}{c|}{PBE+D2}   & \multicolumn{3}{c|}{ACE}      & \multicolumn{3}{c|}{TurboGAP} & \multicolumn{3}{c|}{GAP20} & \multicolumn{3}{c|}{PANNA}     \\ 
    & Gr-ite & Dia & Gr-ene & Grite & Dia & Gr-ene & Gr-ite & Dia & Gr-ene & Gr-ite & Dia & Gr-ene & Gr-ite & Dia & Gr-ene \\ \hline
$C_{11}$ &   1095 $\pm$ 25 &  1042 $\pm$ 8 &    238 $\pm$ 12 &   1045 &  1004 &    275 &    945 &  1024 &    261 &   1022 &   924 &    273 &   1045 &   1054 &    235 \\
$C_{12}$ &    179 $\pm$ 51 &   131 $\pm$ 6 &     39 $\pm$ 25 &    182 &   141 &     44 &    142 &   104 &     39 &    210 &    24 &     47 &    213 &    131 &     58 \\
$C_{13}$ &     -5 $\pm$ 25 &               &                 &     13 &       &        &     12 &       &        &     24 &       &        &     -8 &       &  \\
$C_{33}$ &     58 $\pm$ 10 &               &                 &     27 &       &        &    186 &       &        &    110 &       &        &     33 &       &  \\
$C_{44}$ &      1 $\pm$ 5  &  556 $\pm$ 23 &                 &     9  &   537 &        &  10 &    524 &         &     35 &    474 &        &     -9 &    461 &       \\
$C_{66}$ &    457 $\pm$ 36 &               &     96 $\pm$ 18 &   431  &       &    115 &    402 &       &    111 &    406 &        &    113 &   403 &        &    89 \\ \hline
\end{tabular}
\caption{Elastic moduli of graphite (Gr-ite), graphene (Gr-ene) and diamond (Dia); all valueas are in GPa. }\label{tab:elastic_prop}
\end{table*}


The computed elastic moduli for graphene, diamond and graphite are listed in Table~\ref{tab:elastic_prop} and the phonon spectra plotted along high symmetry directions of the Brillouin zone and the corresponding phonon densities of states are shown in Fig.~\ref{fig:phonons}. The ACE predictions agree closely with the PBE+D2 reference for all three structures. {As pointed out by the PANNA developers}~\cite{Shaidu2021_PaNNA2021}, long wavelength undulations of graphene or graphite sheets are sensitive to numerical details and can induce slightly negative phonon branches close to the $\Gamma$ point as well as slightly negative elastic moduli. Elastic and phonon properties for other bulk structures are provided in the supplementary material~\cite{suppl}. Most of these phases show a number of elastic and phonon instabilities that are accurately captured by ACE but not by TurboGAP, GAP20 or PANNA.

\subsection{Point defects} \label{sec:point_defects}

\begin{figure}
\centering
\includegraphics[width=0.49\textwidth]{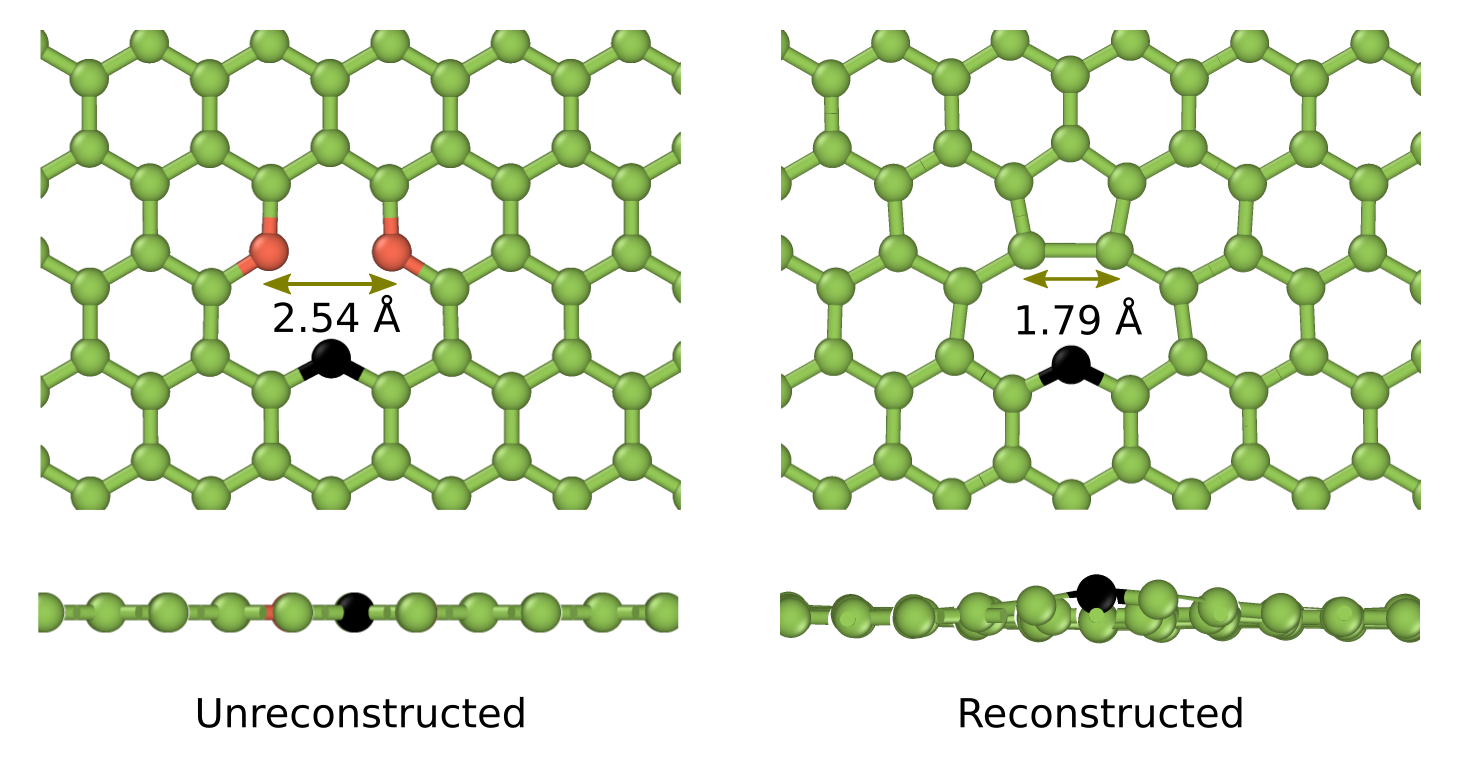}
\caption{Three-fold $D_{3h}$ (left) and reconstructed $C_{2v}$ (right) monovacancy predicted by ACE. Bottom panel: side view of graphene sheet, out-of-plane displacement of the marked atom is about 0.4 \AA. 
}
\label{fig:gra_monovac_restructuring}
\end{figure}

\begin{figure*}
\centering
\includegraphics[width=0.9\textwidth]{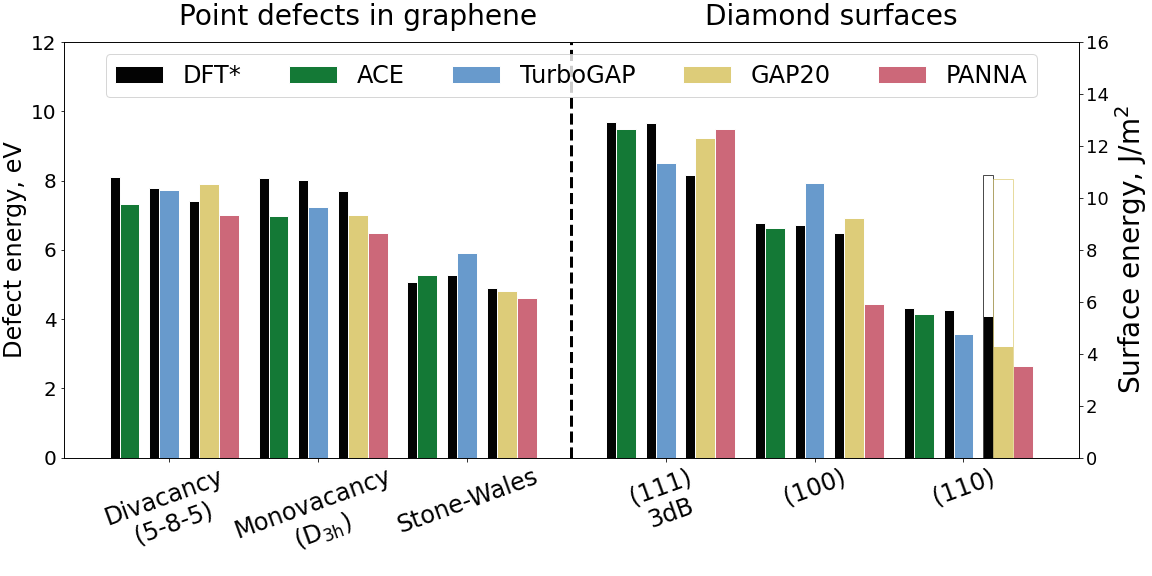}
\caption{Point defects in graphene (left panel) and surfaces in diamond (right panel). See text for details on DFT calculations ; the empty bars for the (110) surface correspond to values reported in the original GAP20 reference~\cite{gap20_doi:10.1063/5.0005084}.}
\label{fig:defects}
\end{figure*}

The directional covalent bonds in carbon generally imply large point defect energies and significant local reconstructions. Vacancies and their clusters influence a broad range of electronic, physical and mechanical properties in graphene~\cite{han2014graphene, lopez2015increasing, skowron2015energetics, zambudio2021, lui2019}. An unreconstructed monovacancy in graphene, formed by removing a single carbon atom, has a three-fold $D_{3h}$ symmetry with three dangling bonds. It undergoes a Jahn-Teller distortion~\cite{skowron2015energetics} and reconstructs to a configuration with lower $C_{2v}$ symmetry consisting of 5- and 9-membered rings. In Fig.~\ref{fig:gra_monovac_restructuring} we show that ACE correctly reproduces this reconstruction, including an out of plane displacement of the central atom. The energy difference between unreconstructed and reconstructed configuration presents an upper bound of the monovacancy migration barrier, predicted to be 0.37 eV by ACE, in close agreement with a range of DFT values~\cite{skowron2015energetics}.
Most classical as well as the considered ML potentials are unable to account for the reconstruction and predict the three-fold symmetric structure as the only stable monovacancy configuration~\cite{qian2021comprehensive}. 

The removal of two neighbouring atoms in graphene allows for a better saturation of dangling bonds. There exist three stable divacancy configurations. The simplest one consists of one 8-membered ring with two adjacent 5-membered rings (5-8-5 configuration). A rotation of a pair of bonds in the 8-member ring results in two further configurations with lower energies, the 555-777 and 5555-6-7777 configurations, with increased numbers of the adjacent rings. {The formation energies of various configurations are reported in Table S3 in the supplementary material}~\cite{suppl}. ACE predicts correctly that the 555-777 divacancy is more stable than the 5-8-5 and 5555-6-7777 configurations by about 0.8 and 0.5 eV, respectively.  Since, the formation energies of divacancies are comparable to that of the monovacancy, it is favorable for two monovacancies to coalesce.

Figure~\ref{fig:defects} (left panel) shows a comparison of computed energies for unreconstructed vacancy defects in graphene. For a fair comparison, the DFT references are those used for the construction of the potentials, namely, PBE+D2 for ACE, PBE+MBD for TurboGAP, and optB88-vdw for GAP20~\cite{gap20_doi:10.1063/5.0005084}. The DFT data for PANNA were not provided, but we assume them to be similar to those for GAP20. It should be noted that the differences between different DFT methods are comparable to the differences between the fitted and reference values. 

\begin{figure}[ht!]
\centering
\includegraphics[width=0.49\textwidth]{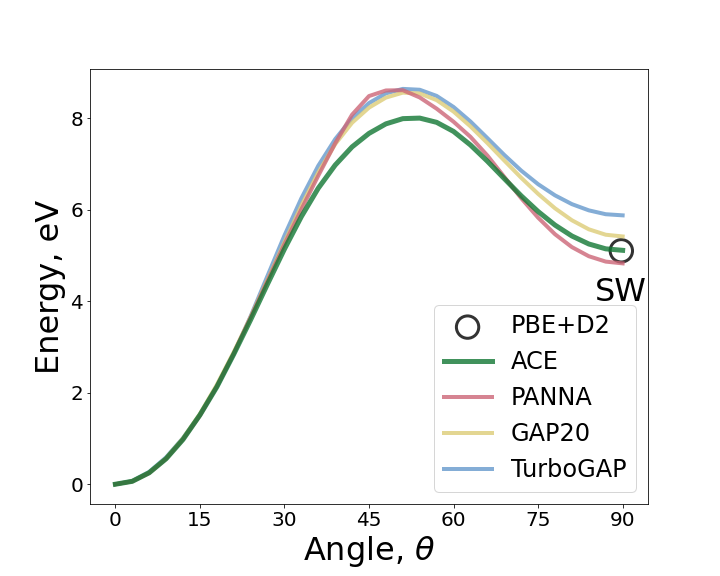}
\caption{Bond rotation for formation of Stone-Wales (SW) defect.}
\label{fig:stone_Wales_rotatation}
\end{figure}

The point defect in graphene with the lowest energy is the Stone-Wales (SW) defect. It is formally generated by a 90$^{\circ}$ rotation of one C-C bond, transforming four 6-member rings into two 5-member rings and two 7-member rings. A consistent description of the SW defect and its formation mechanism implies that a potential models local changes in hybridization correctly and is critical for simulating more complex structures, such as the recently reported monolayer amorphous graphene~\cite{monolayer_a_grapheneToh2020_nature}. We computed the energy barrier associated with C-C bond rotation to be $\sim 8$ eV and the energy of the SW defect as $4.91$ eV, both in excellent agreement with PBE+D2 reference as shown in Fig.~\ref{fig:stone_Wales_rotatation}. 

Overall, ACE predicts structures and energies of basic point defects in graphene in close agreement with the reference DFT values and a comparable level of accuracy is obtained also for other defects~\cite{suppl}. 

\subsection{Diamond surfaces} \label{sec:surfaces}

The right panel of Fig.~\ref{fig:defects} shows the energy of relaxed but unreconstructed low-index {diamond surfaces. 
The empty bars} for the (110) surface correspond to the values reported in the original GAP20 publication~\cite{gap20_doi:10.1063/5.0005084}, which we were unable to reproduce. ACE predicts the energies of all surfaces within 3\% error, while the errors of TurboGAP and GAP20 are larger.

Reconstructions of diamond surfaces driven by relaxation of dangling bonds induce subtle atomic displacement patterns that are a challenging test for interatomic potentials. For example, due to dangling bonds the unreconstructed $(111)$ and $(100)$ surfaces have significantly higher energies than the $(110)$ surface. Two distinct surface terminations, with three (3db) and with one (1db) dangling bonds, exist for the $(111)$ surface. We examined surface reconstructions by relaxing atomic positions to minimize the energy of larger supercells. The simulations were initialized to break the symmetry of the unreconstructed surfaces, either manually or by short MD simulations. 

\begin{figure}
\centering
\includegraphics[width=0.45\textwidth]{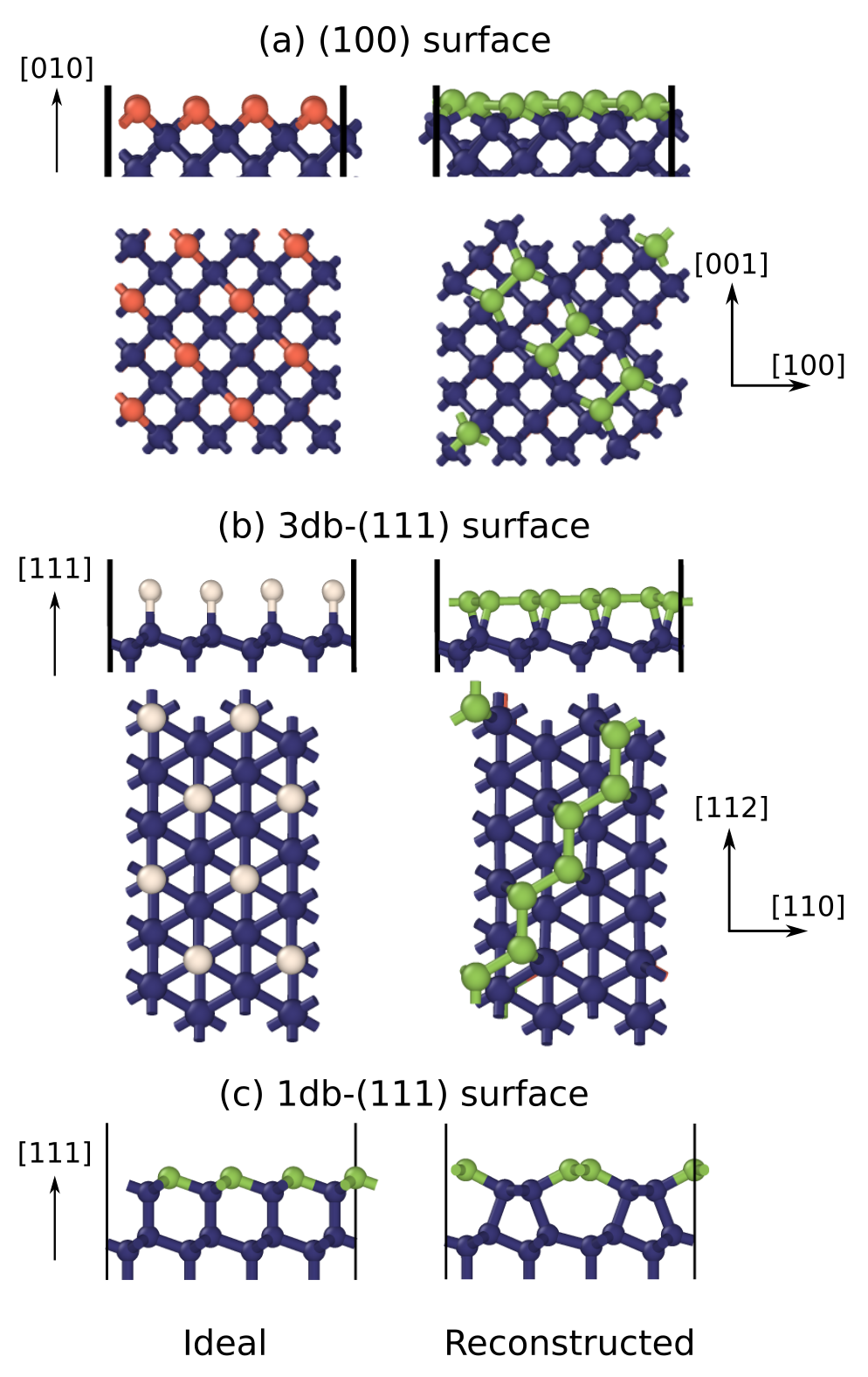}
\caption{Top views of the (a) $(100)$  and  (b,c) $(111)$  surfaces with reconstructions predicted by ACE. White, red, green and blue corresponds to 1-fold, 2-fold, 3-fold and 4-fold coordinated atoms.
}
\label{fig:surface_reconstruction}
\end{figure}

Figure~\ref{fig:surface_reconstruction} shows side and top views of the ideal and reconstructed $(100)$, 3db-$(111)$ and 1db-$(111)$ surfaces. Surface atoms with low coordinations tend to rearrange to achieve a more favourable sp$^2$ (green) hybridization. The 3db-$(111)$ surface undergoes the so-called Pandey-chain reconstruction~\cite{Pandey_reconstruction_PhysRevB.25.4338}, which rearranges surface atoms into $\pi$-bonded chain structures. It is a delicate reconstruction and the surface layers show a strong tendency to graphitize~\cite{chadi_dia_111_surfs}. The 1db-$(111)$ surface instead creates $\pi$-bonded chains between the surface atoms. All reconstructions have a dramatic effect on the surface energies, which are reduced by 2.4, 7.3 and 0.3 J/m$^2$ for the $(100)$, 3db-$(111)$ and 1db-$(111)$ surfaces, respectively. The surface energies are summarized in Table~\ref{tab:surface_energies}.

\begin{table}[]
\tiny
\begin{tabular}{lccccc}
\hline
Surface  & DFT    & ACE                  & TurboGAP             & GAP20           & PANNA          \\ \hline
\bf{(100)}     &        & \multicolumn{1}{l}{} & \multicolumn{1}{l}{} & \multicolumn{1}{l}{} & \multicolumn{1}{l}{} \\
 ~~Ideal       & 9.33 (9.33) & 9.10      & 11.85              & 10.09       & 8.97 	     \\
 ~~Relaxed     & 9.04 (9.08) & 8.78      & 10.25              & 9.13        & 5.92 	      \\
 ~~Reconstr.   & 4.97 (4.83) & 6.08      & 5.45               & 4.96        & 5.92 	      \\ \hline
\bf{3db-(111)} &        & \multicolumn{1}{l}{} & \multicolumn{1}{l}{} & \multicolumn{1}{l}{} & \multicolumn{1}{l}{} \\
 ~~Ideal       & 13.01 (13.02) & 12.72              & 11.37              & 13.45   &   12.81         \\
 ~~Relaxed     & 12.93 (12.98) & 12.71              & 11.37              & 12.17   &   12.65          \\
 ~~Reconstr.   &               & 7.36 	            & 6.56               & 6.24    &    6.72      \\ \hline
\bf{1db-(111)} &        & \multicolumn{1}{l}{} & \multicolumn{1}{l}{} & \multicolumn{1}{l}{} & \multicolumn{1}{l}{} \\
 ~~Ideal       & 7.25 (8.10) & 6.72                & 6.56               & 5.12      &  4.80         \\
 ~~Relaxed     & 5.72 (6.46) & 5.28                & 4.48               & 3.68      &  3.84       \\
 ~~Reconstr.   & 3.51 (3.76) & 4.96 	            & 4.32               & 3.68     &  4.00         \\ \hline
\bf{(110)}     &        & \multicolumn{1}{l}{} & \multicolumn{1}{l}{} & \multicolumn{1}{l}{} & \multicolumn{1}{l}{} \\
 ~~Ideal       & 6.77 (7.46) & 7.13                & 6.88               & 5.76      & 5.12         \\
 ~~Relaxed    & 5.76  (6.50) & 5.73                & 4.80               & 4.16      & 3.52        \\ \hline
\end{tabular}
\caption{Surface energies in J/m$^2$ for ideal, relaxed and reconstructed low-index diamond surfaces. DFT: PBE+D2, in brackets: B3LYP from~\cite{delapierre_2013_doi:10.1080/00268976.2013.829250}}\label{tab:surface_energies}
\end{table}

\begin{figure}
\centering
\includegraphics[width=0.5\textwidth]{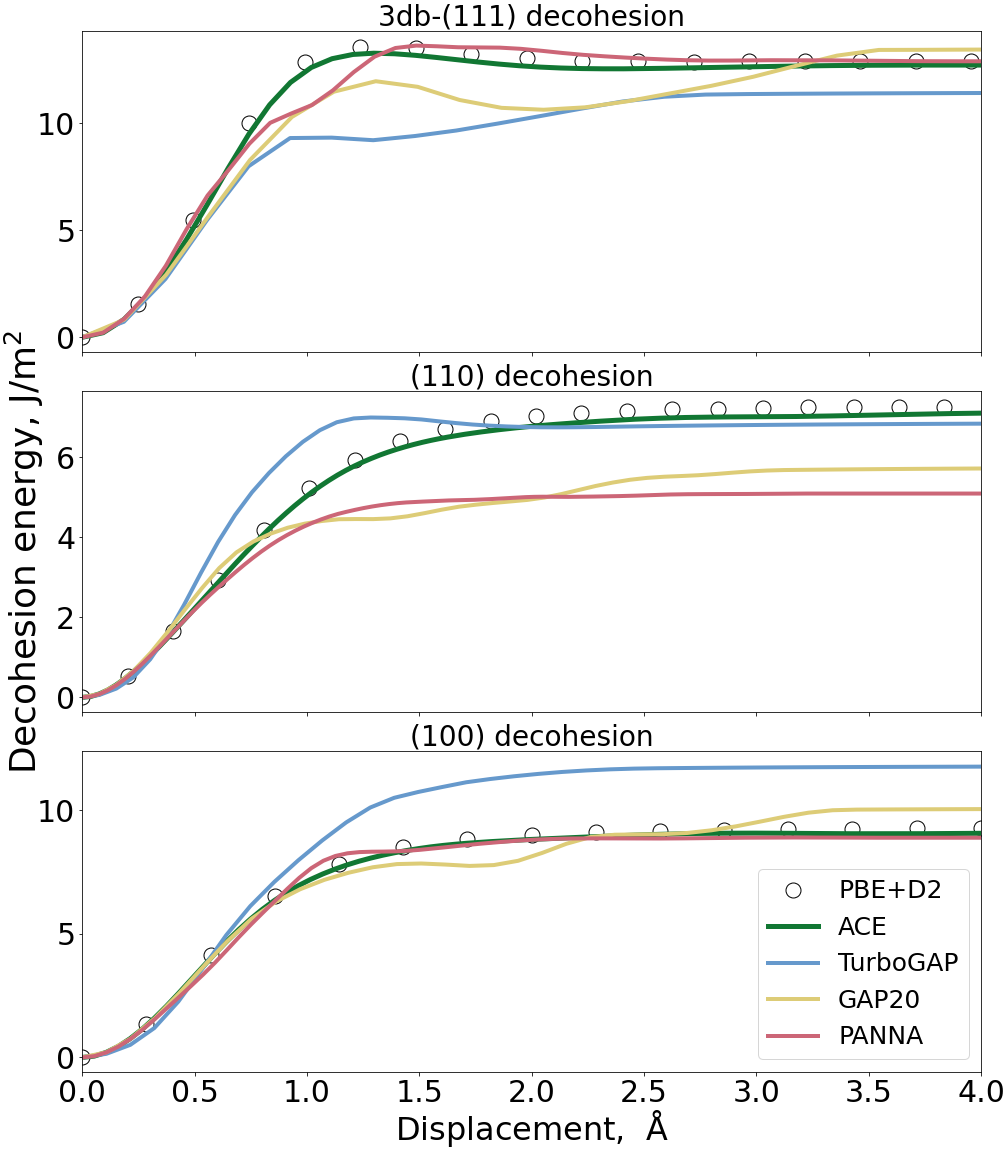}
\caption{Decohesion of unrelaxed, unreconstructed diamond crystals along the $(111)$, $(110)$ and $(100)$ crystallographic orientations. 
}
\label{fig:aitts}
\end{figure}

In addition to static surface calculations, we carried out cleavage simulations to assess smoothness of the energy landscape during bond breaking. A bulk diamond cubic crystal was rigidly separated, excluding relaxations and reconstructions, in the direction normal to the surface using periodic supercells with 32 (for the $(100)$ surface) or 48 (for the $(110)$ and $(111)$ surfaces) atoms. The corresponding variations of the energy as a function of the  separation are shown for all orientations in Fig.~\ref{fig:aitts}. The energy increase is steepest for the $(111)$ surface orientation, for which one set of the C-C bonds is oriented parallel to the loading direction.  DFT (PBE+D2) predicts that the energy plateaus at a separation of  $\approx 2$ \AA{} with a shallow maximum between 1.0 and 1.5 \AA{}. For the $(100)$ and $(110)$ surface orientations, with bonds inclined with respect to the loading direction, the energy increases less steeply and without any barrier.  For the three orientations, ACE captures the PBE+D2 reference closely. TurboGAP and especially GAP20 exhibit oscillations, while PANNA predicts relatively smooth decohesion curves.

\section{Applications}
\label{sec:applications}

\subsection{Brittle fracture of diamond} \label{sec:decohesion_crack}

\begin{figure}
\centering
\includegraphics[width=1\columnwidth]{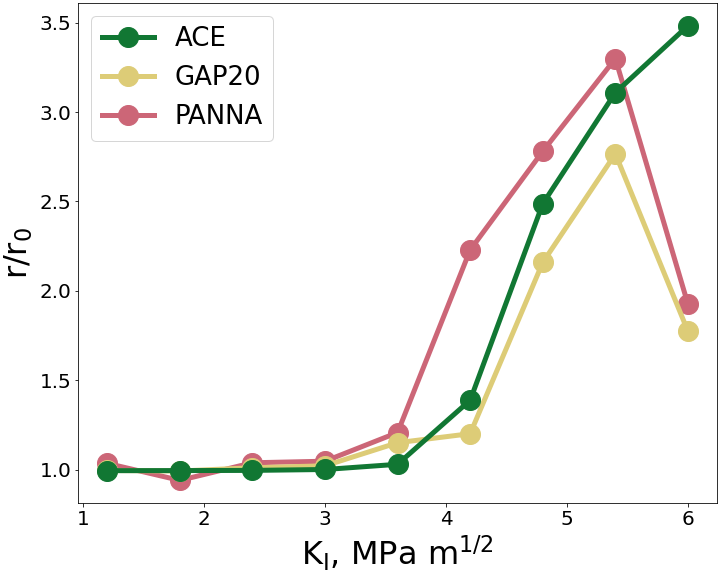}
\caption{Relative distance between atoms at crack tip as function of applied stress intensity factor $K_\text{I}$. }
\label{fig:crack_propagation_plot}
\end{figure}

Fracture simulations of brittle materials are very challenging as they require transferable models which are able to describe bond breaking processes under large and inhomogeneous stresses. At the same time, the models need to remain numerically efficient to be able to simulate large supercells with complex crack geometries at finite temperatures and over realistic time scales~\cite{andric2018atomistic, bitzek2015atomistic}. Here we present MD simulations of brittle cleavage of diamond performed using ACE, GAP20 and PANNA potentials, mostly intended to compare the predictions of these three models. More extended study of diamond fracture will be presented elsewhere.

We simulated semi-infinite cracks with periodic boundary conditions applied along the crack front. The atomic configurations were generated by applying a given stress intensity factor, $K_\text{I}$, to model the asymptotic crack tip region in linear elastic fracture mechanics without applying traction on the outer cell boundaries~\cite{andric2018atomistic} using the code \atomsk~\cite{atomsk_HIREL2015212}.  
Depending on the magnitude of the applied $K_\text{I}$, the crack either tends to heal or to propagate during the simulation.  By varying K$_\text{I}$, one can estimate the critical stress intensity factor, $K_\text{IC}$.

We simulated nine cracks with different magnitudes of the stress intensity factor using LAMMPS \cite{LAMMPS}. The simulation cells {(shown in Fig. S8 in Supplementary material}~\cite{suppl}) contain 5280 atoms with the crack plane normal oriented along the $\langle 011 \rangle$ direction and the $ \langle 111 \rangle$ propagation direction~\cite{suppl}. The initial crack tip was always located in between two $ \langle 011 \rangle$ planes in the center of the simulation cell. We followed the crack evolution for 2 ps using $NVT$ MD simulations at $T=300$ K. The change of the interatomic distance $r$ between two atoms at the initial crack tip  (see the supplementary material~\cite{suppl}) was used to quantify the healing or propagation of the crack. The variation $r/r_0$, where $r_0$ is the initial bond length, as a function of the applied $K_\text{I}$ is plotted for ACE, GAP20 and PANNA in Fig.~\ref{fig:crack_propagation_plot}. We were not able to run equivalent simulations using TurboGAP since it is not implemented in LAMMPS. 

\begin{figure*}
\centering
\includegraphics[width=0.98\textwidth]{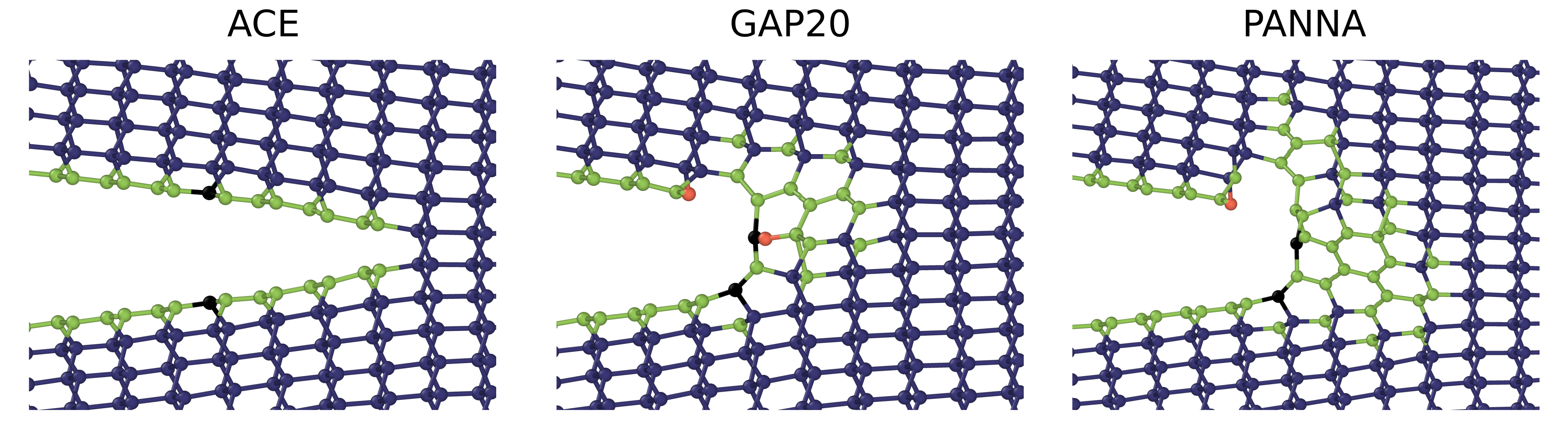}
\caption{Snapshots of crack configurations at $K_\text{I} = 6$ MPa m$^{1/2}$ after 1~ps as predicted by different models; Green color highlights the atoms with sp$^2$ bonding. GAP20 and PANNA show local structural transformations into graphitic structures at the crack tip, while ACE demostrates brittle cleavage.}
\label{fig:compiled_crack_600}
\end{figure*}

Our simulations show that below the critical loading all models predict a closing of the crack along the crack plane. However, ACE is the only model that sustains brittle cleavage, when the loading exceeds the critical value of about 4.2 MPa m$^{1/2}$.  Both GAP20 and PANNA show instead local structural transformations into graphitic structures which lead to blunting of the crack tip, as displayed in Fig.~\ref{fig:compiled_crack_600}. The formation of the graphitic structures explains the sharp drop of $r/r_0$ for both potentials visible in Fig.~\ref{fig:crack_propagation_plot}. To our best knowledge this behavior has not been observed in any theoretical or experimental studies. As discussed in Sec.~\ref{sec:basic_tests}, the phase transformation may be related to the overestimated energy of the diamond phase with respect to those of graphene and graphite due to the DFT reference data employed in the construction of the GAP20 and PANNA potentials.

\subsection{Amorphous carbon}
\label{sec:amorphous_C}

The great variability of amorphous carbon (a-C) networks, governed by competing sp, sp$^2$ and sp$^3$ hybridizations, poses another difficult challenge for atomistic simulations. Two extensive comparative studies of fourteen interatomic potentials~\cite{marks1_transferability, marks2_graphitization} showed that there exist marked differences in the predictions of structural and physical properties of a-C systems. Among the investigated models, GAP17 \cite{gap17_PhysRevB.95.094203} was found to provide the most reliable description (except of some unphysical predictions of 5-fold coordinated atoms at high densities). However, it had by far the highest computational cost of all potentials, limiting its use in large-scale simulations. Even though GAP17 was successfully applied to study the deposition of thin a-C films~\cite{Caro2018_PhysRevLett.120.166101, GAP_application_thinfilm_PhysRevB.102.174201}, larger system sizes and extended simulation times are crucial to achieve realistic amorphous networks. For instance,  Jana et al.~\cite{Jana_2019} investigated in detail the effect of quench rates on the formation and properties of a-C structures using GAP17, the screened Tersoff (Tersoff-S) potential~\cite{Pastewka_SiC_PhysRevB.87.205410} and DFT. The simulations revealed a crucial role of the quench rate on the resulting a-C morphology. Slower cooling rates allowed the atoms to achieve energetically more favourable configurations, thus giving rise to structures with lower cohesive energies, while fast cooling rates resulted in more distorted and less stable forms of a-C.

We studied properties of bulk a-C samples prepared with the liquid-quench MD protocol from Ref.~\cite{Jana_2019}. Simple cubic supercells containing 8000 atoms were melted during 4.0 ps using $NVT$ MD at 12000 K, employing the Nos\'e-Hoover thermostat and time step of 1 fs. The liquid phase was then equilibrated at 8000 K for 10 ps before quenching it to 300 K by linearly decreasing the temperature. We employed three different quench rates: fast at 1000 K/ps, medium at 100 K/ps, and slow at 10 K/ps.  The final structures were optimized by relaxing either the atomic positions only or both the atomic positions and the cell vectors to minimize the stresses in the cells. For the latter protocol, we observed negligible changes of densities in most samples. For each quench rate, we generated ten a-C samples with densities ranging from 1.8 to 3.5 g/cm$^3$. This range encompasses low-density nanoporous structures, crystalline graphite ($\rho_\text{gra} = 2.24$ g/cm$^3$), and diamond ($\rho_\text{dia} = 3.54$ g/cm$^3$).

\begin{figure}
\centering
\includegraphics[width=\columnwidth]{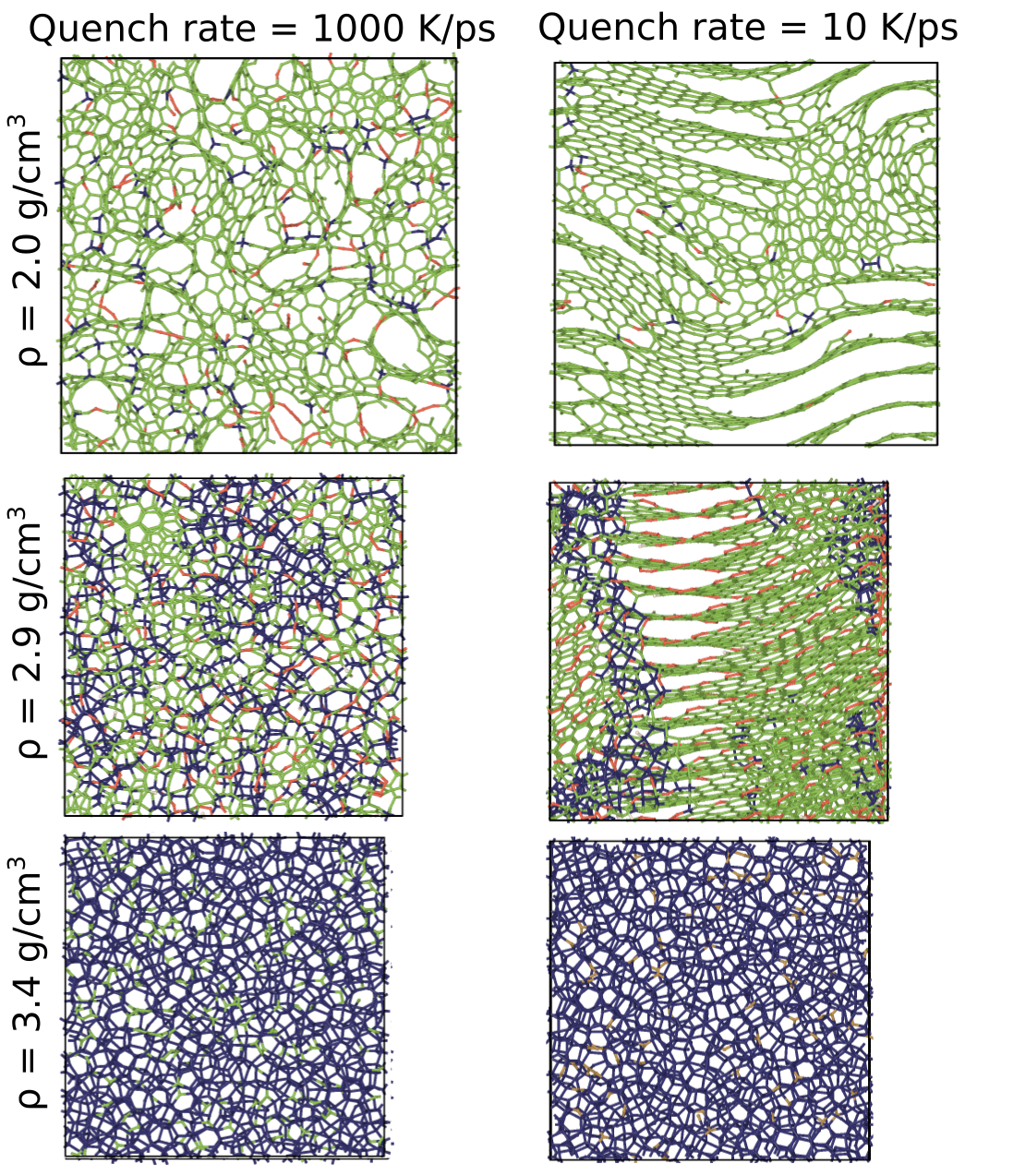}
\caption{Representative a-C structures at three densities using 1000 K/ps (left) and 10 K/ps (right) quench rate. Shown are slices of 1 nm thickness that were cut out of the simulation cell. White, red, green and blue colors correspond to 1-fold, 2-fold, 3-fold and 4-fold coordinated atoms, respectively. Coordination from 1.85 \AA{} cutoff.}
\label{fig:a_C_snapshots}
\end{figure}

Snapshots of representative a-C structures with three different densities generated using the fast and slow quench rates are depicted in Fig.~\ref{fig:a_C_snapshots}. Samples generated with the fast and medium quench rates exhibit uniformly disordered structures that differ in the fraction of the sp$^2$ and sp$^3$ bonded atoms, in agreement with Ref.~\cite{Jana_2019}. The structures of lowest density ($\rho$ = 1.8 g/cm$^3$) are composed of highly distorted and defective graphene sheets with a considerable number of nanovoids and sp-bonded carbon chains connecting the sheets. The structures of intermediate densities ($\rho$ = 2.2 to 2.9 g/cm$^3$) correspond to commonly synthesized a-C structures and the fast and medium quench rates result in disordered glassy networks with a homogeneous mixture of sp$^2$ and sp$^3$ bonded atoms. The structures with the highest density ($\rho$ = 3.4 g/cc) contain mostly sp$^3$ bonded, diamond-like atoms.

In the thermodynamic limit of infinitesimally slow cooling we expect graphite, diamond or coexisting graphite and diamond at relative phase fractions that are determined by density. At the slow cooling rate we observe the onset of phase separation and the occurrence of ordered structures with separated sp$^2$ and sp$^3$ regions. For the lowest density, slow quenching leads to more extended sheets with fewer defects. Such graphitized nanostructures have been observed in previous simulations with GAP17~\cite{marks2_graphitization,Jana_2019}, TurboGAP~\cite{TurboGAP} and experimentally motivated simulations by Bhattarai et al.~\cite{exp_aG_BHATTARAI2018168,exp_aG_C8CP02545B}.

\begin{figure}
\centering
\includegraphics[width=\columnwidth]{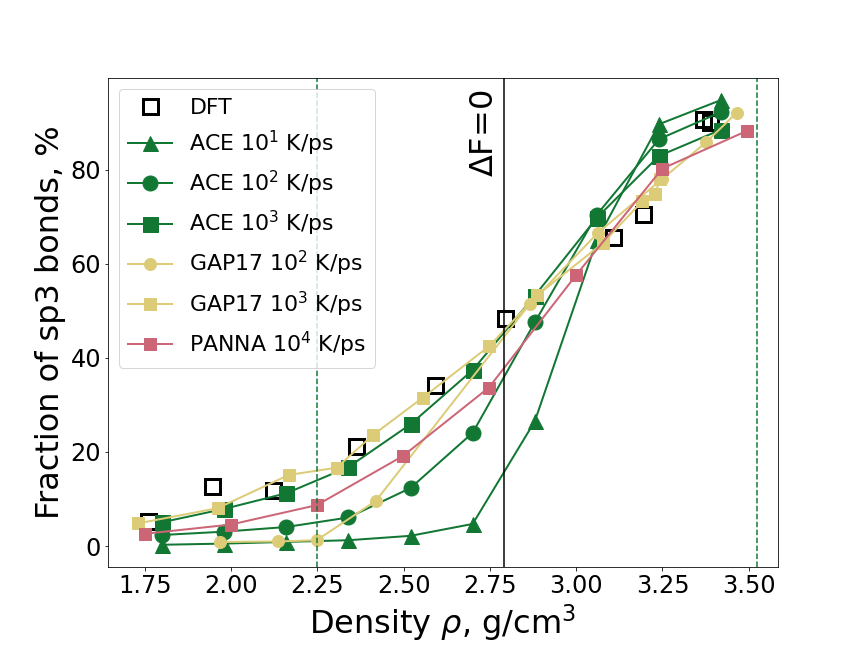}
\caption{Variation of sp$^3$ bond fraction in a-C with density for different quench rates and methods (see text for details). 
}
\label{fig:sp3_fraction}
\end{figure}


Figure~\ref{fig:sp3_fraction} displays how the fraction of sp$^3$ bonded atoms changes as a function of density for different quench rates and potentials. Results for GAP17 (4087 atoms) and DFT (216 atoms) were taken from Ref.~\cite{Jana_2019}, PANNA (216 atoms) was taken from Ref.~\cite{Shaidu2021_PaNNA2021}.  As expected the sp$^3$ fraction increases with density for all methods and quench rates. For the first time we observe a clear influence of quench rate on the a-C morphology with a non-classical potential. For the fast quench at 1000 K/ps, which is also feasible with DFT, the sp$^3$ fraction increases almost linearly with density and leads to a homogeneous amorphous network, see Fig.~\ref{fig:a_C_snapshots}. In contrast, at the slow quench rate of 10 K/ps the sp$^3$ fraction is almost negligible up to $\rho$ = 2.55 g/cm$^3$ and then increases sharply to reach the high values of diamond-like a-C structures. In Fig.~\ref{fig:sp3_fraction}, we mark the equilibrium densities of graphite and diamond by dashed vertical lines. The solid vertical line indicates the density at which the free energies of homogeneously compressed graphite and dilated diamond coincide at $T=0$ K. 






\begin{figure}
  \centering
  \includegraphics[width=\columnwidth]{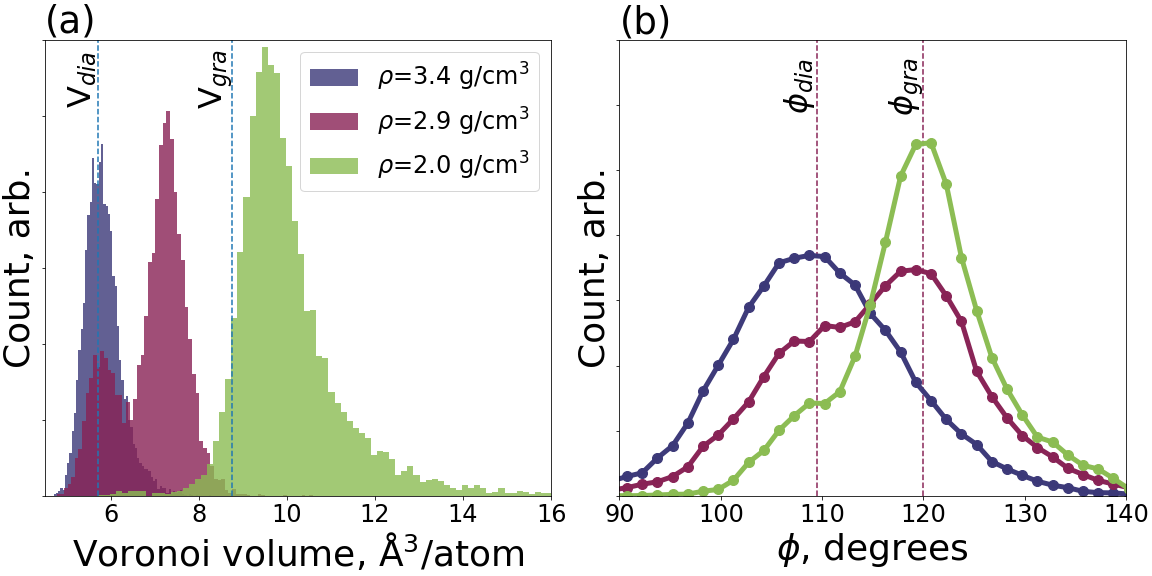}
  \caption{
    Distribution of (a) Voronoi atomic volumes and (b) bond angles in a-C structures at densities 2.0, 2.9 and 3.4 g/cc, quenched at 10 K/ps with ACE.}
  \label{fig:voronoi_adf}
\end{figure}

The distribution of Voronoi atomic volumes, calculated using \pyscal\cite{Menon2019_pyscal}~, for the slowly quenched samples with densities of 2.0, 2.9 and 3.4 g/cm$^3$, are plotted in Fig.~\ref{fig:voronoi_adf} (a). For the highest density sample, there is a single peak (dark blue) centered about the atomic volume of diamond. In contrast, the sample with the intermediate density is characterized by two peaks (dark red). The smaller peak coincides again with the volume of diamond while the higher peak is located in between the diamond and graphite volumes. This results is in accordance with the phase separation into sp$^2$ and sp$^3$ dominated regions observed in Fig.~\ref{fig:a_C_snapshots}. The position of the higher peak indicates that the sp$^2$ phase is a compressed form of amorphous graphite as it is located below the equilibrium volume of crystalline graphite. The distribution of the low density sample (green) is broadest and skewed towards larger volumes due to the existence of nanovoids.

A qualitatively similar outcome can be seen on the distributions of bond angles displayed Fig.~\ref{fig:voronoi_adf} (b). The angular distribution function (ADF) enables to characterize lattice distortions by comparing the positions and widths of the ADF peaks with bond angles in ideal graphite and diamond. The ADFs are centered around $\phi$ = 109.5$^{\circ}$ for the high-density sample and $\phi$ = 120$^{\circ}$ for the low-density sample. For the intermediate-density sample, the distribution is significantly broader and clearly composed of two overlapping peaks centered at the angles mentioned above. The ADFs for structures generated using the medium and high quench rates are shown in {Fig. S10 in the supplementary material}~\cite{suppl}.

\subsection{Fullerene formation}

\begin{figure*}
\centering
    \includegraphics[width=0.99\textwidth]{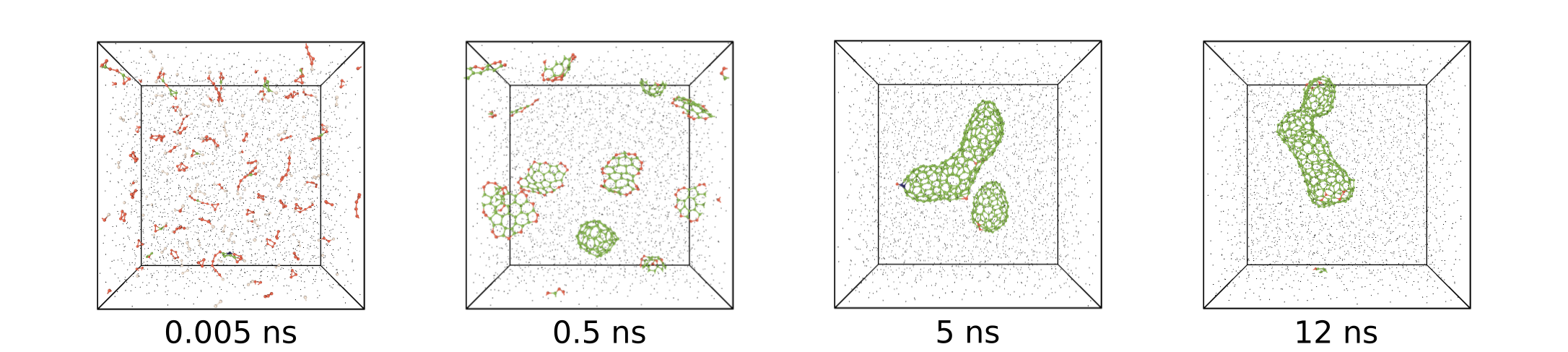}
    \caption{Formation of fullerenes from gas phase during combustion simulation. Small dots are Ar atoms.\label{fig:fullerene_growth}}
\end{figure*}


Carbon clusters are often formed during combustion of carbon-rich materials at high temperatures and pressures. The nucleation and growth process of carbon was the subject of various experimental and theoretical studies\cite{Greiner1988,Millicent2017,Korets2010,Pineau2008,Qin2020,Los2009}.  Here, we present long-time MD simulations of the nucleation and growth of molecular fullerenes from gas phase carbon at high pressure and temperature.

As the starting configuration, in a cubic supercell of side 47.6 \AA{}, 402 C atoms and 2973 Ar atoms were arranged randomly. The Ar-Ar and Ar-C interactions were modeled using a simple Lennard-Jones potential with parameters taken from Refs.~\cite{Pineau2008,Shashkov_1979}, while the carbon interactions were modelled using ACE. The Ar atoms serve as a proxy to exert pressure in the cell and to induce collisions between the C atoms, but do not participate in chemical reactions. The $NVT$ ensemble was used to run MD at 3000 K for 12 nanoseconds. Snapshots from the simulation are shown in Fig.~\ref{fig:fullerene_growth}. {(See Fig.~S11 in the supplementary material}~\cite{suppl} {for snapshots predicted by PANNA, showing clear qualitative deviations from ACE predictions.)}

Similar to the findings of Pineau et al.~\cite{Pineau2008}, {ACE predictions show the gas phase carbon atoms bonding together early in the simulation} to form small buckyball molecules. As the system evolves, the buckyballs interact and coalesce into larger fullerenes. Eventually, at around 12 ns all 402 atoms have merged into a large sp2 bonded fullerene cluster.

\section{Summary and conclusions} 
\label{sec:conclusions}

We developed a general purpose ACE parametrization for carbon that surpasses the accuracy and transferability of state-of-the-art ML models at a fraction of their computational cost. The outstanding predictive power of ACE stems from its physically and chemically motivated formulation, smooth extrapolative properties of the ACE basis, and carefully chosen and internally consistent training data.

We validated the potential extensively through a number of challenging tests. ACE predicts accurately structural and thermodynamic properties for a broad range of ideal and defective carbon polytypes and captures the complex bonding of carbon including bond distortions and bond breaking and making. We showed several exemplary cases where the best available ML potentials fail while ACE predictions are correct.

The efficiency and robustness of ACE was demonstrated on three distinct applications. In simulations of diamond fracture we showed that ACE maintains brittle cleavage when the system is strained beyond the critical load. In contrast, GAP20 and PANNA both predict a graphitic phase transformation at the crack tip. 
Simulations of non-equilibrium amorphous carbon structures reveal that their  structural morphology depends not only on the density but also strongly on the quenching rate. This result could only be achieved due to the outstanding computational efficiency of ACE that enables much slower quenching rates than was possible with other ML potentials.  Lastly, we examined the capability of ACE to describe the  evolution of large fullerene clusters during combustion at high temperatures and pressures. The nucleation and growth of these clusters requires not only long-time MD simulations but also a reliable description of bond formation under highly non-equilibrium conditions.

In summary, the carbon ACE opens new possibilities for structural modeling of carbon at the atomic scale. It not only describes the fundamental properties of carbon allotropes with DFT accuracy, but is also able to maintain this accuracy in large-scale simulations. If necessary, the ACE accuracy and transferability can be further improved systematically, either by tailoring of the training dataset for the required application or by extending the ACE basis. The increased complexity of the ACE parametrization increases the computational costs only linearly, {in contrast to other ML models}~\cite{PACE_Lysogorskiy2021}. 
Finally, the elemental ACE models can be readily extended or combined to address multi-component systems, such as hydrocarbon systems or transition metal carbides.

\section*{Acknowledgements}
The authors acknowledge valuable discussions with Lars Pastewka,  Romain Perriot and Bernd Meyer. MQ acknowledges funding through a scholarship from the International Max Planck Research School for Interface Controlled Materials for Energy Conversion (IMPRS-SurMat).  This work was in part supported by the German Science Foundation (DFG), projects 405621081 and 405621217.

\section*{Supporting Information}
The ACE potential file, training dataset and running examples are included with this work. An additional document~\cite{suppl} including further details of the dispersion corrections, fitting statistics and predictions by other carbon models is also provided. This information is available free of charge via the Internet at https://pubs.acs.org


\bibliography{literature.bib}


\end{document}


\maketitle


\section{Dispersion corrections}
{The base ACE model (termed ACE$_\text{PBE}$) is parametrized to DFT data computed with the PBE functional which is not able to capture the long-range dispersion interactions. As a result, ACE$_\text{PBE}$ is unable to describe binding in graphite. For a transferable carbon potential, dispersion interactions need to be taken into account and this can be achieved by additive corrections that are applied on top of ACE$_\text{PBE}$.}

\begin{figure}
\centering
    \includegraphics[width=0.45\columnwidth]{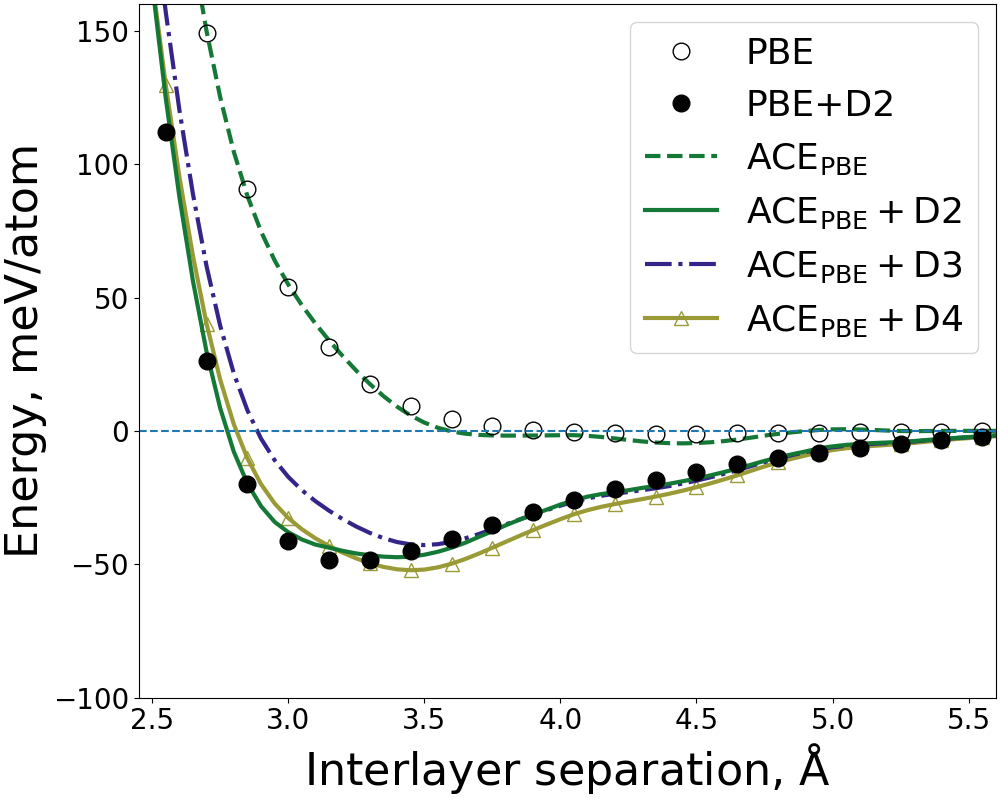}
    \hfill
    \includegraphics[width=0.45\columnwidth]{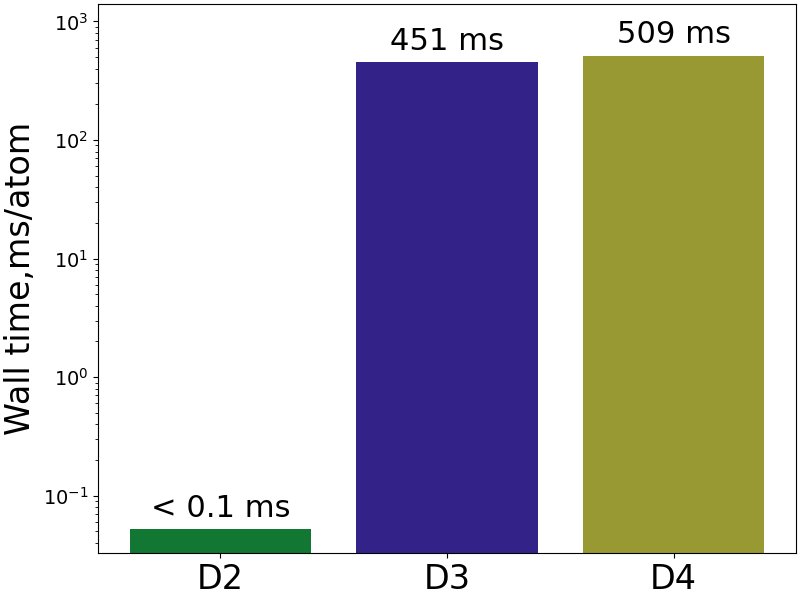}
     \caption{{Binding of graphite as a function of interlayer separation, as predicted by different dispersion corrections to ACE$_\text{PBE}$ (a), and the computational time for each method (b).}\label{fig:dx_comparison} }
\end{figure}

{A number of such corrections for DFT have been proposed, most notably the D2, D3 and D4 corrections by Grimme and co-workers}~\cite{vdw_d2_coorection,D3_Grimme_2010,D3_Grimme_2011,vdw_d4_coorection,Caldeweyher_grimme_D4_2019,Caldeweyher_Grimme_D4_2020}. Figure~\ref{fig:dx_comparison} {shows a comparison between the different correction methods for the binding between graphite layers as a function of the inter-layer separation. As seen in Fig.}~\ref{fig:dx_comparison}{(a), the binding energies of graphite predicted by the D2, D3 and D4 corrections are in close agreement (within $\pm$5 meV/atom). However, since the D3 and D4 correction include environment dependent terms, they are considerably more expensive to evaluate than the pairwise D2 correction, as shown in Fig.}~\ref{fig:dx_comparison}(b).

{Considering the significant difference in computational costs and the negligible differences to reproduce basic dispersion interactions, we employed the simplest D2 correction in the present study (see below for details). However, it is noted that it can be readily replaced by more elaborate correction schemes if necessary (e.g., for studies of friction between graphene layers)}.

\subsection*{D2 Correction}
The D2 term in ACE is expressed as a simple pair potential

\begin{equation}
    E_{disp} = -\frac{1}{2} \sum_{i} \sum_{j} \frac{C_6}{r_{ij}^6} f_{d}(r_{ij})  ,
\end{equation}
where the damping function $f_d$ is given by
\begin{equation}
    f_d = \frac{s_6}{1+\exp(-d (r_{ij}/r_c - 1))} .
\end{equation}
The dispersion coefficient $C_{6}$, the damping factor $d$, and the vdW radius $r_c$ were parameterized for all elements up to Xe~\cite{vdw_d2_coorection}. The global scaling parameter $s_{6}$ depends on the specific DFT functional used and is set to 0.75 for PBE.

\section{Validation of neural network models}

This section includes additional validation of the recent NN models hNN\_Gr$_x$ and DeepMD.

\begin{figure}
\centering
    \includegraphics[width=0.45\columnwidth]{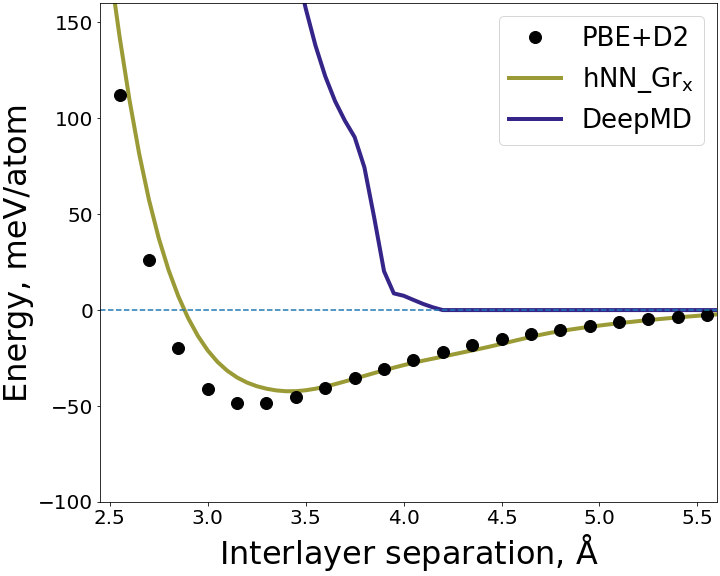}
    \hfill
    \includegraphics[width=0.45\columnwidth]{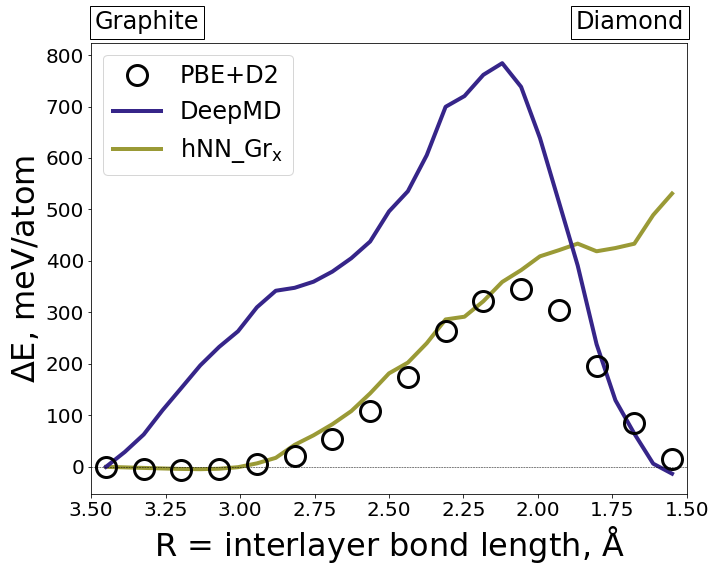}
     \caption{Binding of graphite as a function of interlayer separation (a), and the transformation path from graphite to diamond (b), as predicted by hNN\_Gr$_x$ and DeepMD potentials.\label{fig:nnp_graphite_eb} }
\end{figure}


\begin{figure}
\centering
\includegraphics[width=0.75\textwidth]{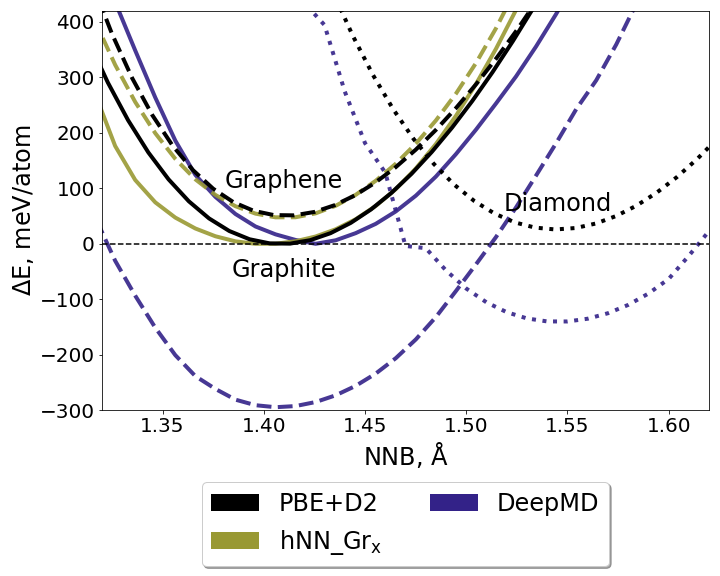}
\caption{Relative stability of the three lowest energy phases of carbon, as predicted by DFT and NN-based models. $\Delta$E refers to the energy difference with respect to graphite given by the respective potential. Solid, dashed and dotted lines refer to graphite, graphene and diamond phase respectively.}
\label{fig:ev_groundstate}
\end{figure}

\clearpage
\section{Comparison of RMSE values for energies and forces}

\begin{figure*}
\centering
\includegraphics[height=0.9\textheight]{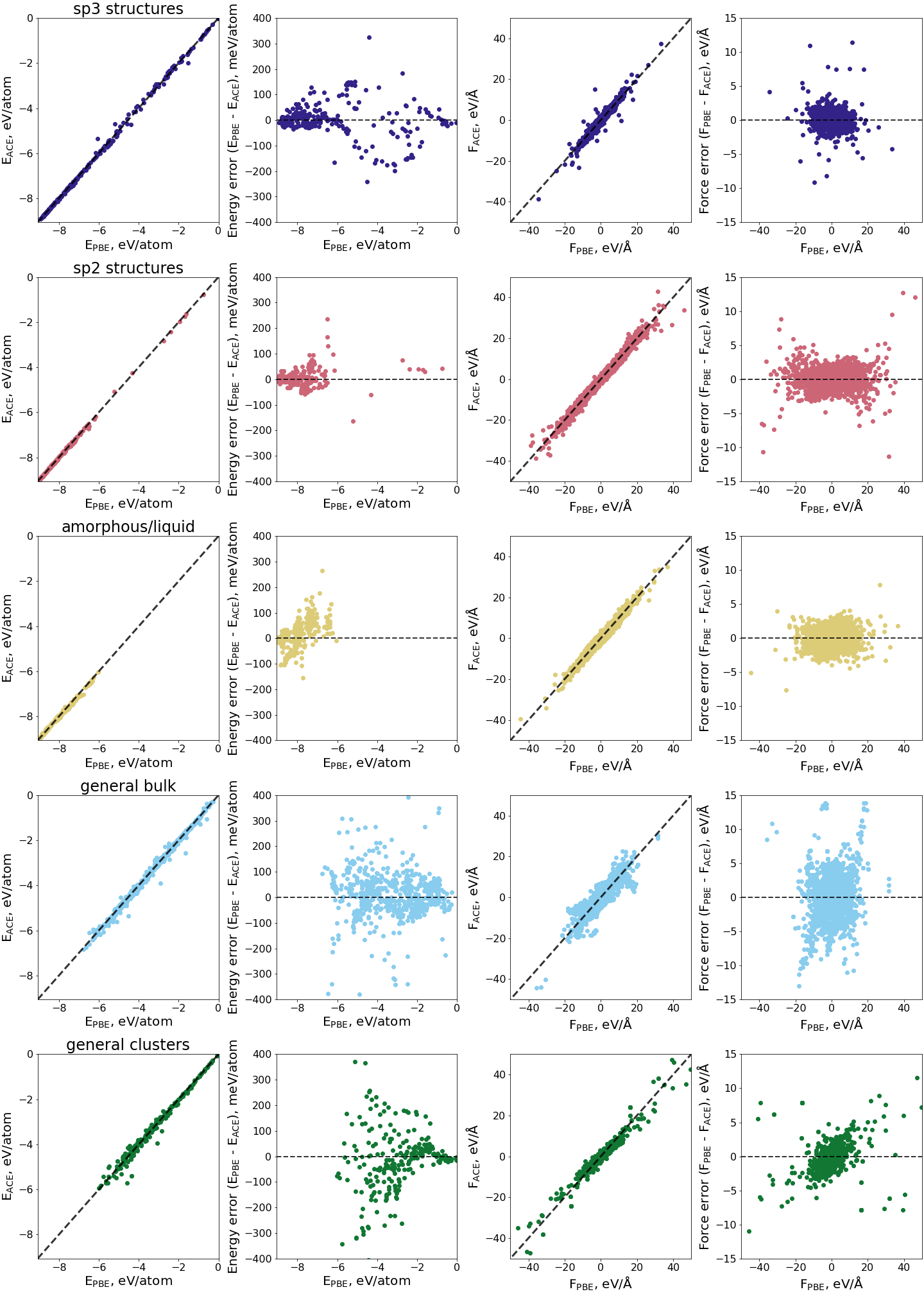}
\caption{Fit metrics for 10\% split test datasets corresponding to different categories. }
\label{fig:all_crossvalidations}
\end{figure*}

\begin{table}[h]
\caption{Comparison of RMSE values for energies and forces for the considered models computed from a test dataset. RMSE values are calculated with respect to PBE data. Energies are expressed in terms of relative cohesive energy with respect to ground state of the given potential to account for different references.}\label{tab:s_rmse_pots}
\begin{tabular}{lcc}
\hline
 Potenial    & E$_\text{RMSE}$ (meV/atom) & F$_\text{RMSE}$ (meV/\AA{}) \\ \hline
ACE      & 102   & 539   \\
TurboGAP & 238   & 883   \\
DeepMD   & 841   & 942  \\
GAP20    & 1083  & 725   \\
hNN\_Gr$_\text{x}$ & 1552  & 1982  \\
PANNA    & 1793  & 1327 \\ \hline
\end{tabular}
\end{table}

\begin{table}
\caption[]{Test RMSEs for a larger ACE fitted with 1950 functions, as indicated by the final point in Figure 3 in the main paper. Uniform weights are applied over all the structures. Numbers in brackets correspond to the number of structures for each category, respectively. Energies are in meV/atom and forces in meV/\AA{}. \label{tab:rmse_dataset}}
\begin{tabular}{lcc}
\hline
    Category (test size) & $E_\text{test}$ & $F_\text{test}$  \\ \hline
   sp$^2$ structures (385) &        23 &   284   \\
   sp$^3$ structures (378) &        57 &   229   \\
   amorphous/liquid (297) &        50 &   542    \\
   general bulk (606)      &        49 &  492   \\
   general clusters (246)  &       161 &  666   \\ \hline
\end{tabular}
\end{table}

\begin{figure}
\centering
    
    \begin{tabular}{@{}c@{}}
    
    \includegraphics[width=0.9\columnwidth]{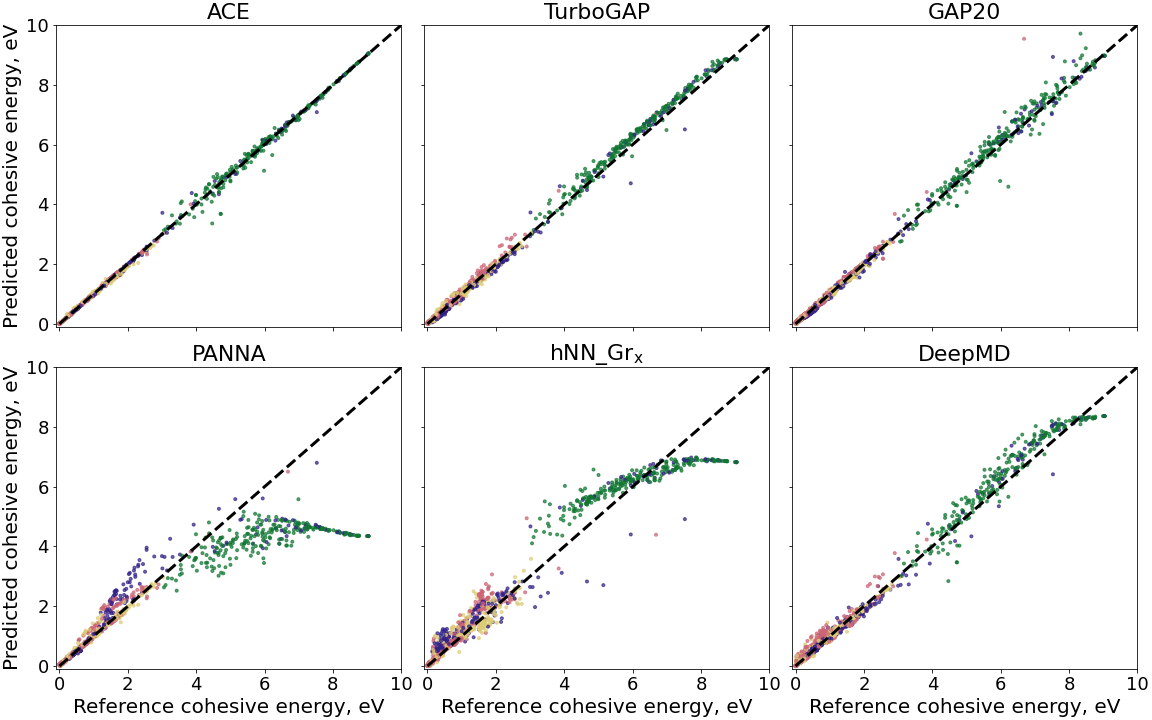}
  \end{tabular}
  
  \small (a)
  
  \vspace{\floatsep}

  \begin{tabular}{@{}c@{}}
    \includegraphics[width=0.9\columnwidth]{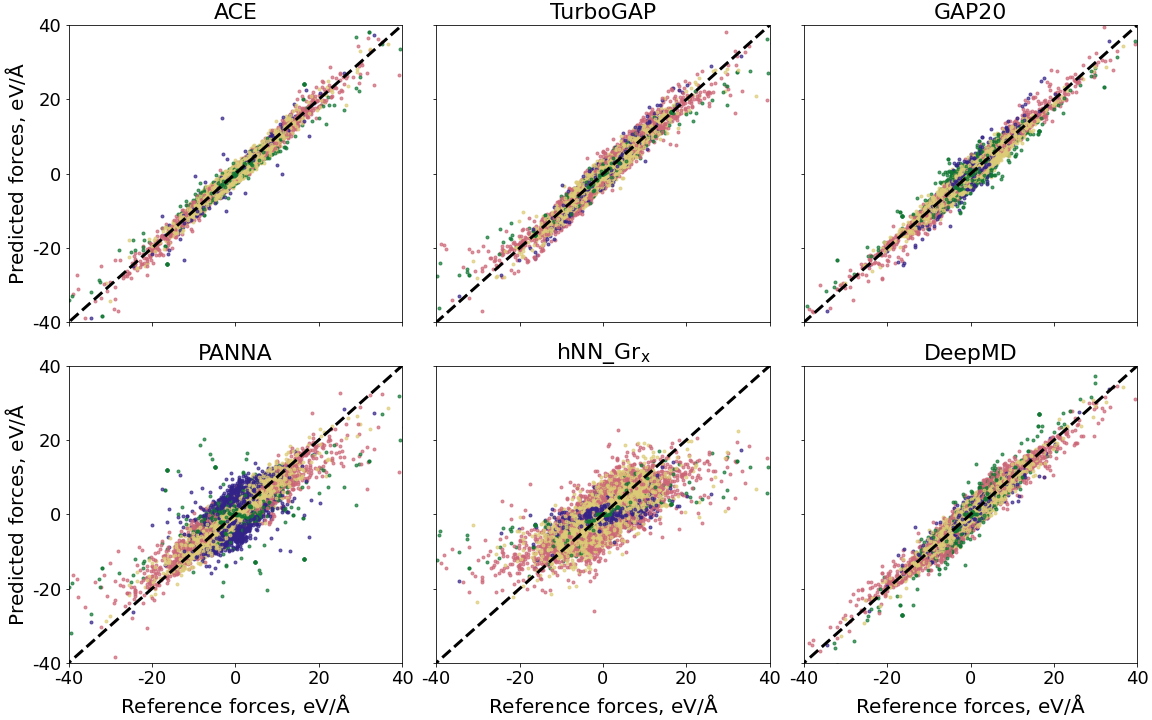} 
  \end{tabular}
  
  \small (b)

    \caption{\label{fig:s_pots_crossref} (a) Cohesive energy and (b) force validation plots for considered models for a split test dataset with 1912 structures. RMSE values are given in Table~\ref{tab:s_rmse_pots}. Points are colored according to the legend given in Fig. 1 in the main paper.}
\end{figure}

\clearpage
\section{Validation of additional models and properties}\label{fig:all_Ev}


\begin{table}[h]
\caption{Formation energies for point defects in graphene and diamond lattice }
\begin{tabular}{lcccccc}
\hline
                        & DFT* & ACE  & GAP20 & PANNA & hNN\_Gr$_\text{x}$ & DeepMD \\ \hline
Graphene                &     &      &     &     &     & \\ ine
Stone-Wales             & 4.9 & 4.9  & 4.8 & 4.6 & 5.3 & 4.8  \\
Monovacancy             & 7.7 & 5.8  & 7.0 & 6.5 & 7.9 & 6.2 \\
Divacancy (5-8-5)       & 7.4 & 7.2  & 7.9 & 7.0 & 8.0 & 7.2 \\
Divacancy (555-777)     & 6.6 & 6.4  & 6.9 & 6.0 & 6.9 & 6.5 \\
Divacancy (5555-6-7777) & 6.9 & 6.9  & 7.4 & 6.3 & 7.3 & 8.0 \\
Adatom                  & 6.4 & 6.4  & 5.9 & 10.8 & 8.3 & 6.7 \\ \hline
Diamond                 &     &      &     &     &    & \\ \hline
Monovacancy             & 6.6 & 5.8  & 4.3 & 5.0 &    & 5.7 \\
Divacancy               & 9.1 & 12.0 & 6.6 & 11.0 &   & 11.6\\ \hline
\end{tabular}

{*DFT reference is taken from~\cite{gap20_doi:10.1063/5.0005084} and is given for optB88-vdW DFT functional}
\end{table}


\begin{figure}

\centering
    \subfloat{\includegraphics[width=0.33\textwidth]{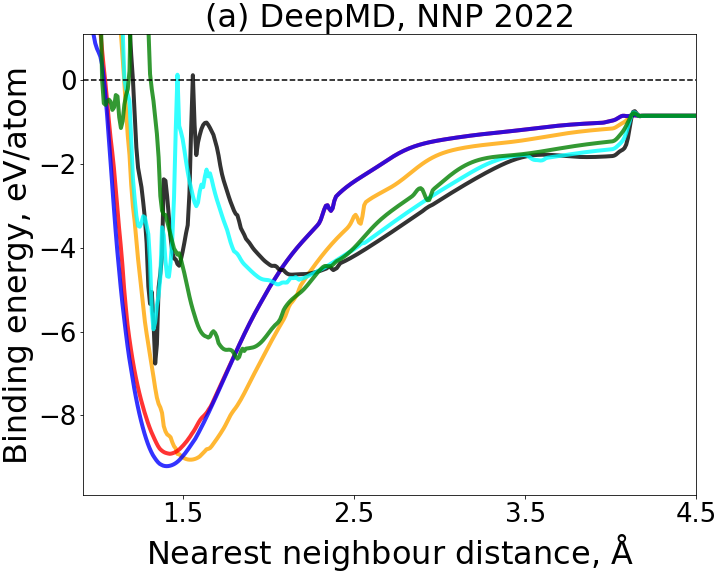}} 
    \subfloat{\includegraphics[width=0.33\textwidth]{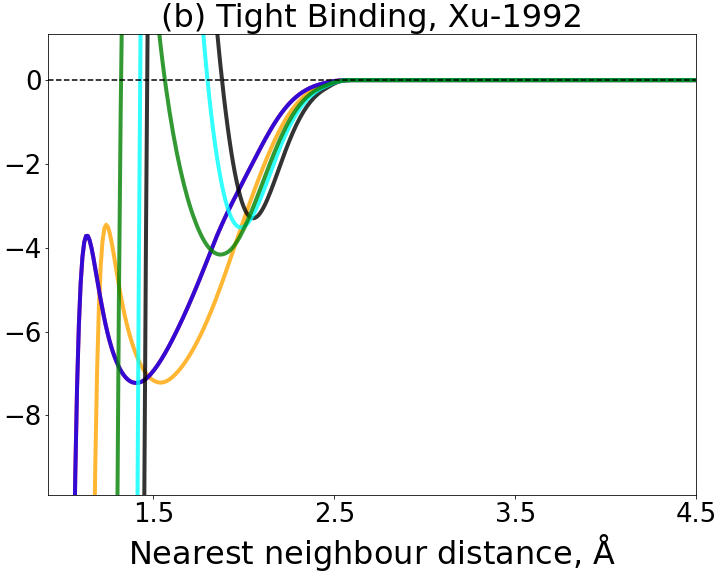}}
    \subfloat{\includegraphics[width=0.33\textwidth]{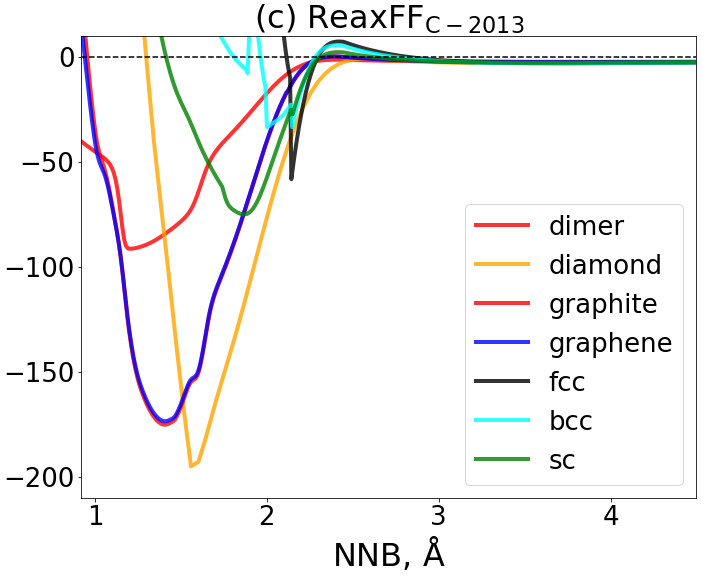}}

    \subfloat{\includegraphics[width=0.33\textwidth]{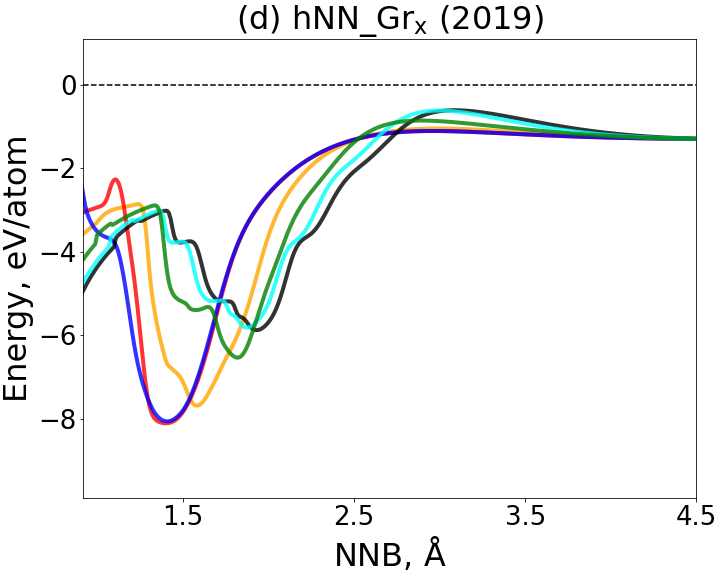}} 
    \subfloat{\includegraphics[width=0.33\textwidth]{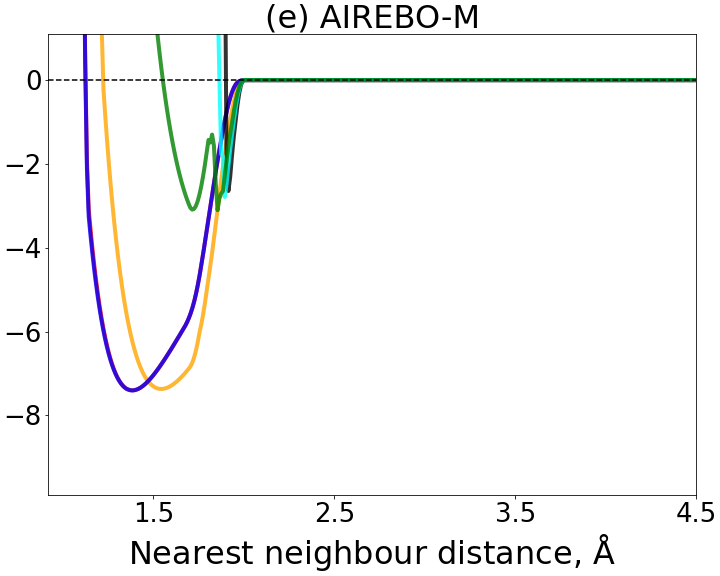}}
    \subfloat{\includegraphics[width=0.33\textwidth]{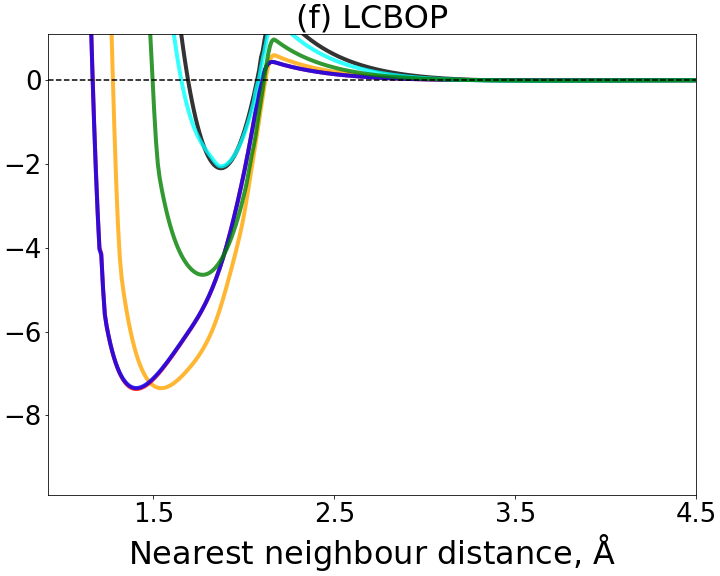}}
    
\caption{Energy volume curves for some bulk phases as predicted by (a) DeepMD NNP~\cite{deep_MD_carbonpot_WANG20221}, (b) Tight Binding model parametrized by Xu et al.~\cite{Xu_1992}, (c) ReaxFF for hydrocarbon oxidation~\cite{reaxff_CHO_Chenoweth2008}. (d) NNP for multilayered graphene~\cite{NNP_Gr_PhysRevB.100.195419}, (e) AIREBO with Morse modification~\cite{AIREBO_M_doi:10.1063/1.4905549} and (f) LCBOP potentials~\cite{LCBOP_PhysRevB.68.024107} \label{fig:ev_suppl} }    
\end{figure}

\begin{figure}
\centering
    \includegraphics[height=0.9\textheight]{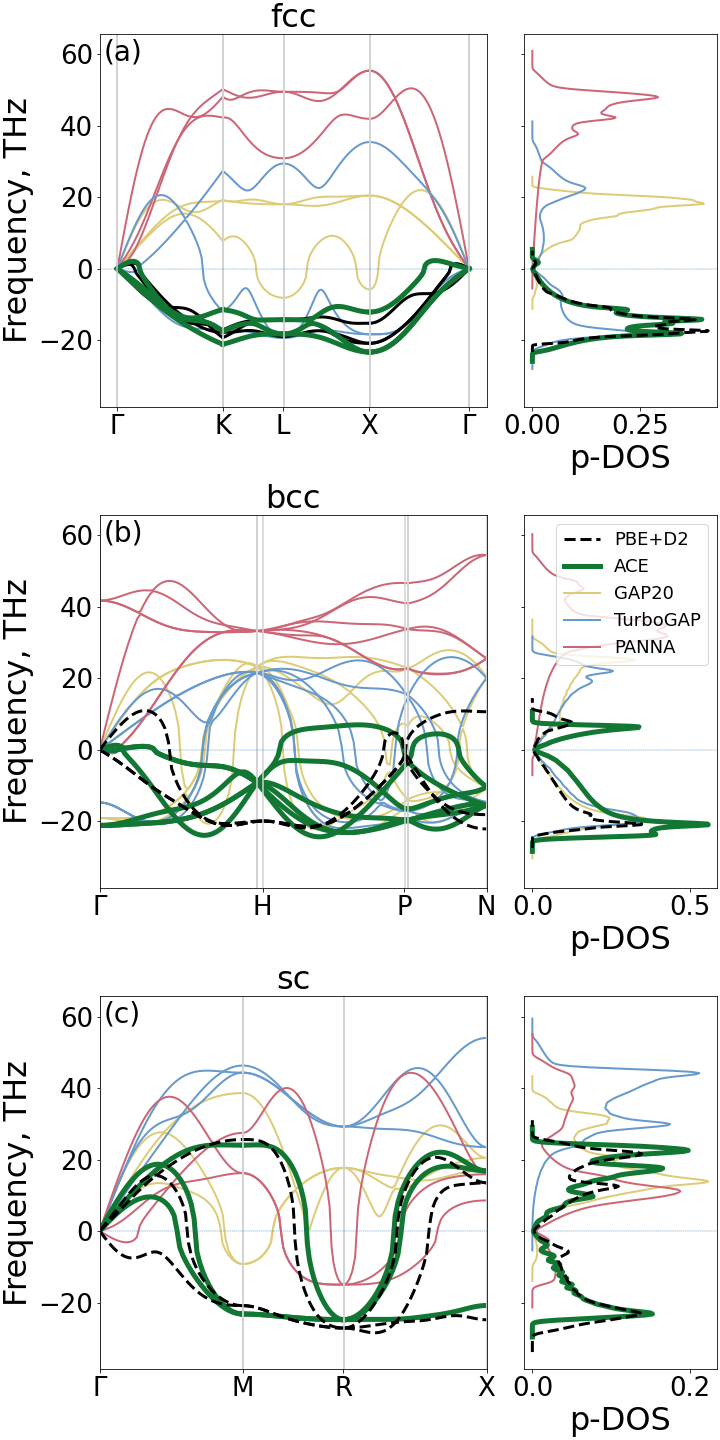}
    \caption{Computed phonon band structures for fcc, bcc and sc bulk phases; all three phases are dynamically unstable according to DFT.\label{fig:other_bulk_phonons} }
\end{figure}

\clearpage
\section{Diamond fracture}

\begin{figure}[h]
  \centering
  \begin{tabular}{@{}c@{}}
    \includegraphics[width=0.5\linewidth]{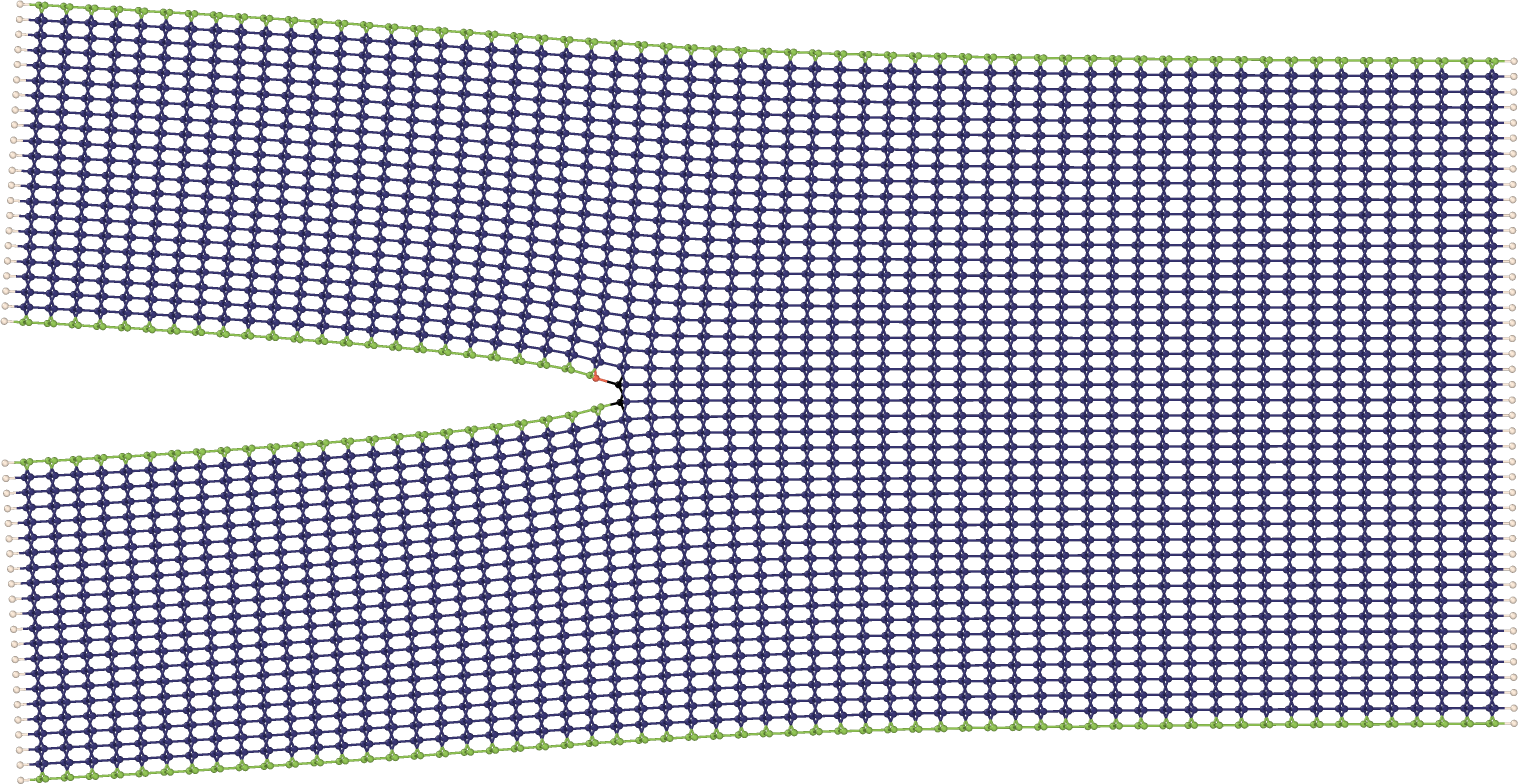} \\[\abovecaptionskip]
    \small (a)
  \end{tabular}

  \vspace{\floatsep}

  \begin{tabular}{@{}c@{}}
    \includegraphics[width=0.5\linewidth]{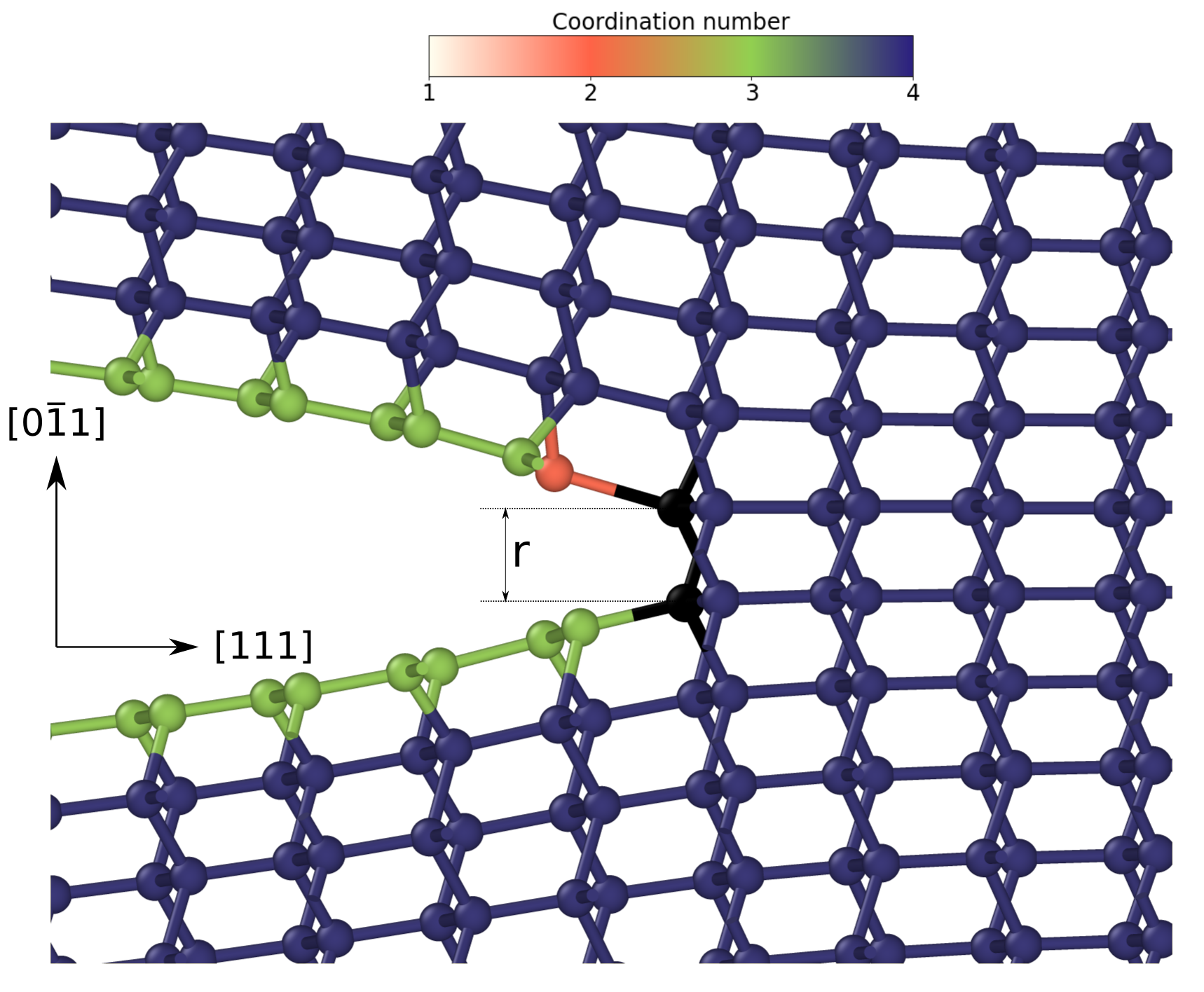} \\[\abovecaptionskip]
    \small (b)
  \end{tabular}

  \caption{An example of simulation cell consisting of 5280 atoms used to study the brittle crack propagation in diamond (a) and a close-up of the crack tip region (b).}\label{fig:crack_tip}
\end{figure}

\clearpage
\section{Amorphous carbon}




\begin{figure}
\centering
    \includegraphics[width=0.5\columnwidth]{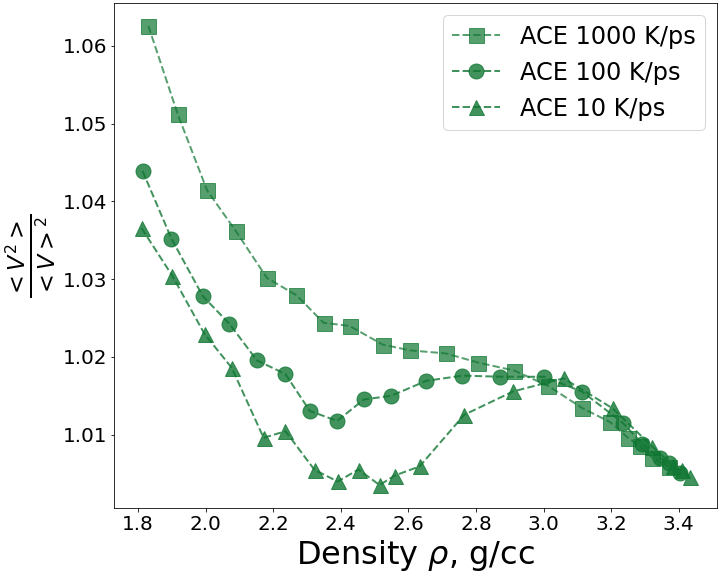}

    \caption{Width of Voronoi volume distribution as a function of density for all generated a-C structures\label{fig:aC_vorwidth} }
\end{figure}


\begin{figure}[h]
\centering
    \includegraphics[width=0.9\textwidth]{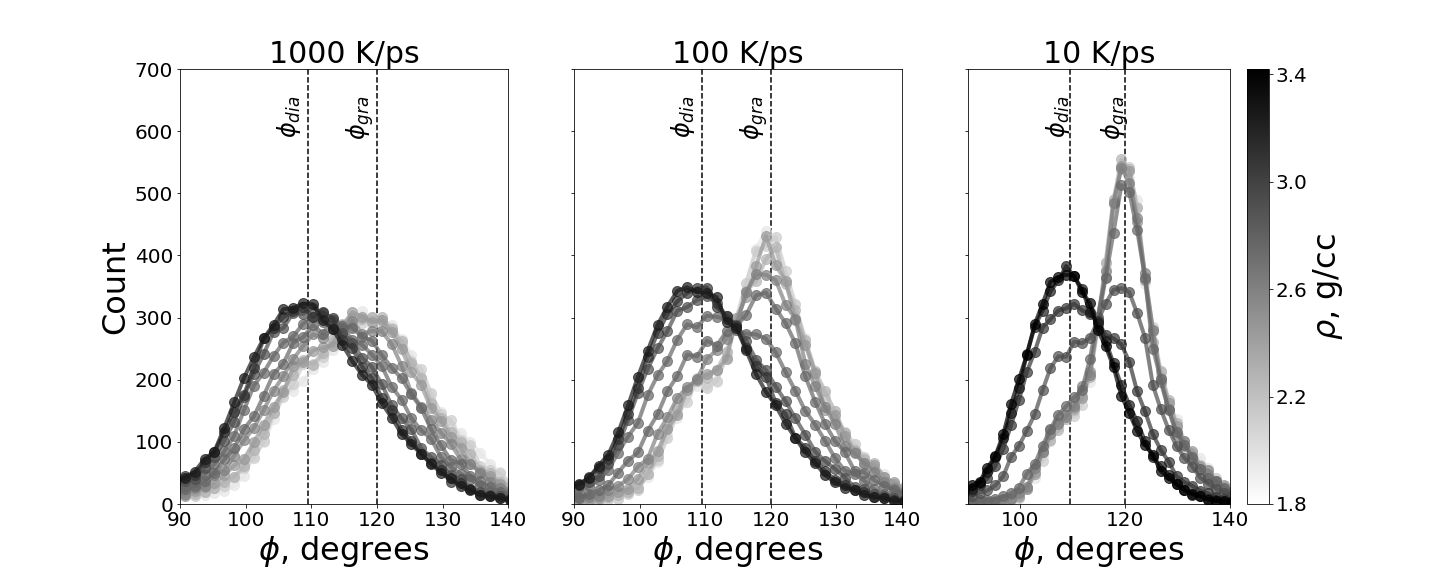}
    \caption{A comparison of angular distribution functions for a-C structures at different quench rates\label{fig:all_adfs} }
\end{figure}

\clearpage
\section{Fullerene formation}
\begin{figure}
\centering
    \includegraphics[width=0.8\columnwidth]{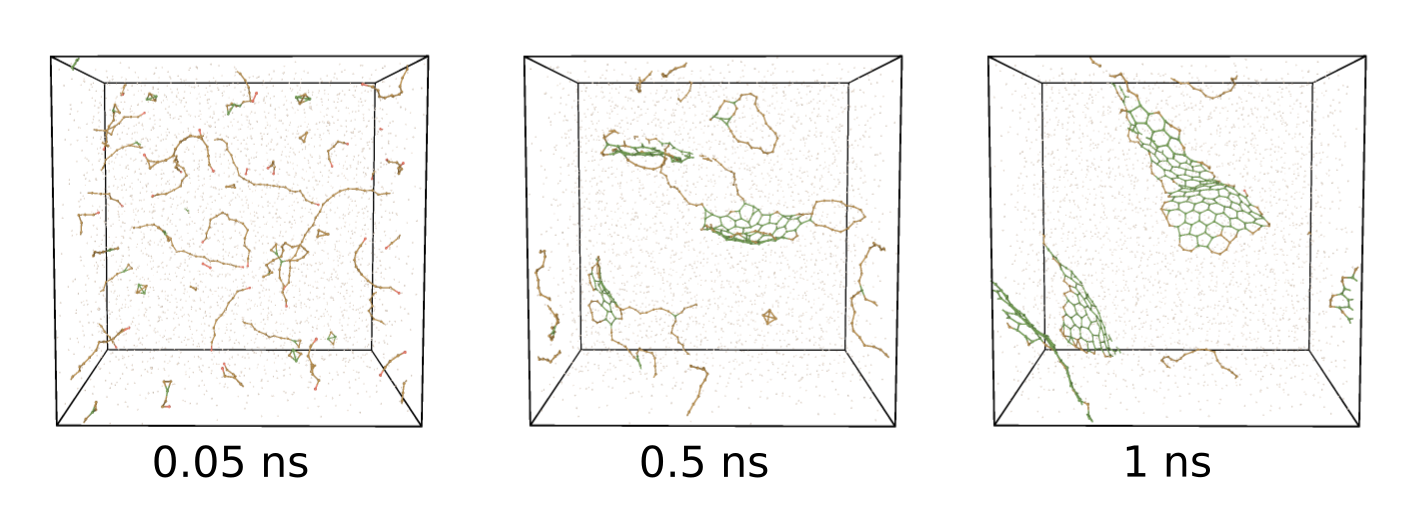}
    \caption{{Formation of fullerenes from gas phase during combustion simulation as predicted by PANNA. In contrast to ACE, PANNA predicts formation of only defective graphitic flakes up to 1 ns rather than fullerene molecules.} \label{fig:panna_fullerene_growth}}
\end{figure}

\clearpage
\section{{Data availability}}

{To help prospective users getting started with the ACE potential, some files are provided:}

\begin{itemize}
    \item \textbf{ACE potential:} {The ACE potential file \textit{`c\_ace.yace'} along with a lammps implementation of D2 \textit{`d2.table' }is provided in the directory \textit{`potential\_file'}}.
    
    \item \textbf{LAMMPS example:} {The directory \textit{`c\_pace\_eq\_melt'} contains an input lammps script to run an example simulation of melting of a C supercell. The directory also contains a \textit{`README'} file explaining the compilation of lammps and use of the potential, both with python and with lammps.}
    
    \item \textbf{Training dataset:} {The training dataset used to parametrize the ACE is provided as a \textit{`pckl.gzip'} file. It can be readily used with the provided} \PM {input file \textit{`fit\_c\_ACE.yaml'} to parametrize a new ACE potential. The training dataset can be analyzed using the python script \textit{`read\_datasey.py'}. More information on the dataset format and} \PM {can be found at https://github.com/ICAMS/python-ace}
    
    \item \textbf{Dispersion correction:} {A jupyter notebook \textit{`dispersion\_corrections.ipynp`} is included which shows how to include different vdW corrections such as D2, D3 and D4 with the ACE potential.}
\end{itemize}

\bibliography{literature.bib}